\documentclass[review]{elsarticle}

\usepackage{hyperref}

\usepackage[utf8]{inputenc}

\usepackage{amsmath}
\usepackage{amsfonts}
\usepackage{amssymb}

\begin{document}

\begin{frontmatter}

\title{Reduced description method in the kinetic theory of Brownian motion with active fluctuations}

\author[addr1,addr2]{Yu.~V.~Slyusarenko\corref{cauthor}}
\cortext[cauthor]{Corresponding author}
\ead{slusarenko@kipt.kharkov.ua}

\author[addr1]{O.~Yu.~Sliusarenko}
\author[addr1,addr3]{A.~V.~Chechkin}

\address[addr1]{Akhiezer Institute for Theoretical Physics National Science Center ``Kharkiv Institute of Physics and Technology'', 1~Akademichna Str., 61108 Kharkiv, Ukraine}
\address[addr2]{Karazin National University, 4~Svobody Sq., 61077 Kharkiv, Ukraine}
\address[addr3]{Department of Physics and Astronomy University of Padova, via Marzolo 8, 35131 Padova, Italy}

\begin{abstract}

We develop a microscopic approach to the kinetic theory of many-particle systems with dissipative and potential interactions in presence of active fluctuations. The approach is based on a generalization of Bogolyubov--Peletminsky reduced description method applied to the systems of many active particles. It is shown that the microscopic approach developed allows to construct the kinetic theory of two- and three-dimensional systems of active particles in presence of non-linear friction (dissipative interaction) and an external random field with active fluctuations. The kinetic equations for these systems in case of a weak interaction between the particles (both potential and dissipative) and low-intensity active fluctuations are obtained. We demonstrate particular cases in which the derived kinetic equations have solutions that match the results known 
in the literature. It is shown that the display of the head-tail asymmetry and self-propelling even in the case of a linear friction, is one of the consequences of the local nature of the active fluctuations.

\end{abstract}


\end{frontmatter}

\section{Introduction}

Active matter is a field in soft matter physics, which studies the properties of aggregates of self-propelled objects that have the ability to take up energy from the environment, to store it in an internal depot, and to convert internal energy into kinetic energy. Active matter comprises diverse systems spanning from macroscopic (e.g. schools of fish and flocks of birds) to microscopic scales (e.g. migrating cells, motile bacteria and gels formed through the interaction of nanoscale molecular motors with cytoskeletal filaments within cells). Here we refer the reader to the reviews \cite{ramaswamy2010mechanics,vicsek2012collective,romanczuk2012active,marchetti2013hyd} and references therein. In recent years the number of publications devoted to theoretical and experimental studies of the phenomena in the active substances has sharply increased, and the recent achievements have been mirrored in the Special Issues of the European Physical Journal Special Topics and the Journal of Statistical Physics \cite{epjst2014,epjst2015,epjst2015_2,epjst2016,jst2015}. The study of non-equilibrium processes in systems of active particles inevitably raises the question of the consistent derivation of the evolution equations for such systems, in particular, the kinetic equations. The kinetic theory of the systems of active particles is a challenging issue that attracts attention during the recent decade. Thus, Bertin et al. \cite{bertin2006boltzmann,bertin2009hydrodynamic,bertin2015comparison} has derived the Boltzmann equation for the self-propelled point-like particles on a two-dimensional plane with the assumption that the modulus of the velocity vector is fixed and identical for all the particles, so that only the direction of the vector plays a role in the dynamics. Ihle \cite{ihle2011kinetic,ihle2014towards} developed an alternative kinetic approach that is based on the Chapman-Kolmogorov equation for the N-particle probability density. The resulting mean-field kinetic equation has been studied analytically and numerically, and extended to the so-called topological interactions \cite{ihle2013invasion,chou2012kinetic,romensky2014tricritical}. Romanczuk et al. \cite{romanczuk2011brownian,grossmann2012active,grossmann2013self} derived and explored the mean-field kinetic equation in two spatial dimensions, starting from the Langevin equation with active friction and active fluctuations, supplemented with different forces describing interaction between the particles. In Ref. \cite{lobaskin2013collective} the authors also pursue the Langevin approach to study collective dynamics in two-dimensional system of active Brownian particles with dissipative interactions.

In our paper we develop a consistent microscopic approach based on the Hamilton equations that take into account the external random forces acting on the particles. The microscopic approach to the construction of the kinetic theory essentially provides dynamic justification of statistical mechanics of many particle systems \cite{bogoliubov1959problems,akhiezer1981methods}. N.N. Bogolyubov suggested a method of reduced description of the evolution of many-particle systems \cite{bogoliubov1959problems}, which allowed the construction of a regular procedure for obtaining closed dissipative kinetic equations based on the BBGKY chain of reversible equations for many-particle distribution functions. Fundamentals of reduced description method were formulated in \cite{bogoliubov1959problems} for the classical (non-quantum) systems of many particles. In case of quantum many-particle systems the ideas of the Bogolyubov reduced description method were developed in the works by S.V.\~{}Peletminsky, and the main results are presented in \cite{akhiezer1981methods}. There are also other approaches for the dynamical justification of statistical mechanics, which are different from Bogolyubov's approach, for example, in the works of Prigogine's Brussels school \cite{prigogine1962}, as well as different formulations of Bogolyubov's ideas, see, e.g., \cite{zubarev1974, klimontovich1986stat, zubarev1996, luzzi2002predictive}. In the present paper we use the reduced description method in the form close to the one of Bogolyubov--Peletminsky \cite{peletminskii2003kinetics}, to construct the kinetic theory of many-particle systems with active fluctuations and non-linear friction. For that purpose we need to generalize the canonical Bogolyubov--Peletminsky approach in order to take into account an external stochastic impact and dissipative interactions. 

A generalization of the Bogolyubov reduced description method to the case of dissipative many-particle systems in an external stochastic field was first suggested in \cite{sliusarenko2015bogolyubov}. In this paper, the authors proposed a formalism for deriving kinetic equations. As a starting point, a stochastic Liouville equation obtained from Hamilton's equations taking dissipation and stochastic perturbations into account was used. The Liouville equation is then averaged over realizations of the stochastic field by an extension of the Furutsu--Novikov formula to the case of a non-Gaussian field. As the result, a generalization of the classical Bogolyubov--Born--Green--Kirkwood--Yvon  hierarchy is derived. In order to get a kinetic equation for the one-particle distribution function, the authors use a regular breaking procedure of the BBGKY hierarchy by assuming weak interaction between the particles and weak intensity of the field. Within this approximation they get the corresponding Fokker--Planck equation for the system in a non-Gaussian stochastic field. Two particular cases by assuming either Gaussian statistics of external perturbation or homogeneity of the system are discussed. In that approach, however, the stochastic external forces do not depend on the velocity (or momentum) of the particles. In other words, the formalism developed in \cite{sliusarenko2015bogolyubov} can be applied to the systems with 
non-linear friction, as it is the case of the active particle systems, but with passive fluctuations
of either Gaussian or non-Gaussian nature.

In the present paper, we suggest a generalized formulation of the reduced description method, suitable for describing the kinetics of many-particle dissipative systems with active fluctuations. It is shown that in the framework of the microscopic approach developed it is possible to construct the kinetic theory of active particles both in the cases of two- and three-dimensional systems, with the availability of non-linear friction (dissipative interaction), as well as local impact of an external active random field. Under the ``local impact'' we assume that this field may act differently at different points in space. In other words, the effect of this field on a particle may depend not only on the velocity (or momentum) of that specific particle, but on the point in the coordinate space where the particle is located. The general kinetic equations for such systems are obtained. We also consider special cases in which the obtained kinetic equations give solutions known for the active particles from the earlier works \cite{romanczuk2012active,romanczuk2011brownian,grossmann2013self,lobaskin2013collective}.

\section{Basics}

Consider a system consisting of $ N $ identical active particles of mass $ m $, each of which is characterized by spatial coordinates $ \mathbf x_{\alpha} $, $ 1 \le \alpha \le N $, measured from the center of mass, and momentum $ \mathbf p_{\alpha} $, $ 1 \le \alpha \le N $. The interaction between the particles is assumed to consist of two parts - a ``reversible'' part described by the Hamiltonian $ H $, and ``irreversible'' one, described by the function $ R^\mathrm{\omega} $, the meaning of which will be explained below.

The Hamiltonian of the system can be written as:

\begin{equation} \label{GrindEQ__1_1_}
H=H_{0} +V=\sum_{1\le \alpha \le N}\frac{\mathbf p_{\alpha }^{2} }{2m}  +\sum_{1\le \alpha <\beta \le N}V_{\alpha ,\beta },
\end{equation}
where $V_{\alpha ,\beta } $ is the pair interaction potential,
\begin{equation} \label{GrindEQ__1_2_}
\begin{gathered}
V_{\alpha ,\beta } \equiv V\left(\mathbf x_{\alpha \beta } \right), \\
\mathbf x_{\alpha \beta } \equiv \mathbf x_{\alpha } -\mathbf x_{\beta }
\end{gathered}
\end{equation}

We also assume that the particles of the system are exposed to specific forces that depend on the particle velocity (or momentum) and are characterized by a function $ R $. We assume that the function $ R $ can be represented as:
\begin{equation} \label{GrindEQ__1_3_}
R=R^{r} +R^\mathrm{\omega },
\end{equation}
where $R^{r} $ is a regular part of this function.
\begin{eqnarray}\label{GrindEQ__1_4_}
\nonumber R^{r} \equiv \sum_{1\le \alpha <\beta \le N}R_{\alpha ,\beta }  ,\\
R_{\alpha ,\beta } \equiv R\left(\mathbf x_{\alpha \beta } ,\mathbf p_{\alpha \beta } \right),\\
\nonumber \mathbf p_{\alpha \beta } \equiv \mathbf p_{\alpha } -\mathbf p_{\beta }  
\end{eqnarray}
and $R^\mathrm{\omega } $ is a stochastic part of the function $R$, which can be written as
\[R^\mathrm{\omega } \equiv \sum_{1\le \alpha \le N}R^\mathrm{\omega } \left(x_{\alpha } ,t\right) , \, x_{\alpha } \equiv \left\{\mathbf x_{\alpha } ,\mathbf p_{\alpha } \right\}.\] 

The stochastic nature of the function $ R^\mathrm{\omega} $ is formally highlighted by the presence of index $ \mathrm{\omega} $.

Note that in the case of non-active identical particles with the dissipative interaction function, $ R^{r} $ is treated as a dissipative function, see \cite{landau1980statisticheskaia} and \cite{sliusarenko2015bogolyubov, goldhirsch2013application}. It is usually assumed that the dissipation in the system is related to friction of macroscopic particles, so that in this case the dissipation function $ R_{\alpha, \beta} $, following \cite{landau1980statisticheskaia}, can be chosen as:
\begin{equation} \label{GrindEQ__1_5_}
\begin{gathered}
R_{\alpha ,\beta } \equiv \frac{1}{2} \tilde{\gamma }\left(\mathbf x_{\alpha \beta } \right)\mathbf p_{\alpha \beta }^{2},\\
\tilde{\gamma }\left(\mathbf x_{\alpha \beta } \right)>0, \\
\mathbf p_{\alpha \beta } \equiv \mathbf p_{\alpha } -\mathbf p_{\beta } 
\end{gathered}
\end{equation}

This implies that $ \tilde {\gamma} \left (\mathbf x_{\alpha \beta} \right) = 0 $ if $ \left | \mathbf x_{\alpha \beta} \right | \mathop {>} \limits_{\sim} r_{0} $, where $ r_{0} $ is a characteristic range of dissipative forces. In view of the property \eqref {GrindEQ__1_5_} the friction coefficient is always positive.

However, in case of the active particles the positivity does not always hold \cite{romanczuk2012active} . The ``friction coefficient'' in the Langevin--type equations for active particles can depend on the velocity and change its sign. Therefore, one can not use the criteria \eqref {GrindEQ__1_5_} to determine the properties of ``dissipative function'' $ R_{\alpha, \beta} $ in the case of active particles. For that reason we use here the quotation marks which however, will be omitted in what follows.

Following the usual classical theoretical mechanics procedures,  and taking into account the Eqs.~\eqref {GrindEQ__1_1_} -- \eqref {GrindEQ__1_4_}, the generalized Hamilton equations for the system under study can be written as 
\begin{equation} \label{GrindEQ__1_6_}
\begin{gathered}
\dot{\mathbf p}_{\alpha } =-\frac{\partial H}{\partial \mathbf x_{\alpha } } -\frac{\partial R}{\partial \mathbf  p_{\alpha } }, \\
\dot{ \mathbf x}_{\alpha } =\frac{\partial H}{\partial  \mathbf p_{\alpha } }.
\end{gathered}
\end{equation}

Thus, the force $ \mathbf  F_{\alpha, \beta} $ acting on a particle $ \alpha $ from the particle $ \beta $  consists of two terms:
\begin{equation} \label{GrindEQ__1_7_}
 \mathbf F_{\alpha ,\beta } = \mathbf F_{\alpha ,\beta }^{p} +F_{\alpha ,\beta }^{r}, 
\end{equation}
namely, the force $ \mathbf  F_{\alpha, \beta}^{p} $, connected with the presence of a potential pair interaction between the particles, and the force $ F_{\alpha, \beta}^{r} $, connected with the presence of a dissipative interaction between the particles (in the sense outlined above)
\begin{equation} \label{GrindEQ__1_8_}
\begin{gathered}
 \mathbf F_{\alpha ,\beta }^{p} \equiv -\frac{\partial V_{\alpha ,\beta } }{\partial \mathbf  x_{\alpha } },\\
 \mathbf F_{\alpha ,\beta }^{r} \equiv -\frac{\partial R_{\alpha ,\beta } }{\partial  \mathbf p_{\alpha } }. 
\end{gathered}
\end{equation}

In addition, it follows from Eq.~\eqref {GrindEQ__1_1_}, that the $ \alpha $ -th particle is influenced by external random force $  \mathbf Y_{\alpha}^\mathrm{\omega} $, which depends on the momentum of the particle, wherein:
\begin{equation} \label{GrindEQ__1_9_}
\begin{gathered}
-\frac{\partial R\left(x_{\alpha } ,t\right)}{\partial p_{i\alpha } } \equiv Y_{i}^\mathrm{\omega } \left(x_{\alpha } ,t\right)=X^\mathrm{\omega } \left( \mathbf x_{\alpha } ,t\right)e_{\alpha i}^{h} +X_{j}^\mathrm{\omega } \left( \mathbf x_{\alpha } ,t\right)\left(\delta_{ij} -e_{\alpha i}^{h} e_{\alpha j}^{h} \right), \\
e_{\alpha i}^{h} \equiv \frac{p_{\alpha i} }{\left| \mathbf p_{\alpha } \right|}.
\end{gathered}
\end{equation}

The latter expression requires some comments. We emphasize, first of all, that the stochastic force $ Y_{i}^\mathrm{\omega} \left (x_{\alpha}, t \right) $ in Eq.~\eqref {GrindEQ__1_9_} is written in a form that is not related to the choice of a particular coordinate system. This notation simply reflects the fact that the stochastic force acts differently along and across the direction of a particle velocity. The expression \eqref {GrindEQ__1_9_} looks like a natural extension of the stochastic force $ Y_{i}^\mathrm{\omega} \left ( \mathbf x, t \right) $ typical for the Langevin equation in case of an ordinary Brownian particle:
\[\dot{ \mathbf x}= \mathbf v, \quad \dot{v}_{i} =-\gamma v_{i} +Y_{i}^\mathrm{\omega } \left( \mathbf x,t\right).\] 
In fact, the value $ Y_{i}^\mathrm{\omega} \left ( \mathbf x , t \right) $ in the last equation can always be identically rewritten as:
\[Y_{i}^\mathrm{\omega } =\left( \mathbf e^{h}  \mathbf Y^\mathrm{\omega } \right)e_{i}^{h} +Y_{j}^\mathrm{\omega } \left(\delta_{ij} -e_{i}^{h} e_{j}^{h} \right),\] 
where $ e_{ih} $ is arbitrary unit vector, for example, $ e_{i}^h = p_i/| \mathbf p| $. Replacing here the scalar product $  \mathbf e^{h}  \mathbf Y^\mathrm{\omega} $ with $ X^\mathrm{\omega} \left ( \mathbf x_{\alpha}, t \right) $, $ Y_{j}^\mathrm{ \omega} $ by $ X_{j}^\mathrm{\omega} \left ( \mathbf x_{\alpha}, t \right) $ and assuming $ e_{\alpha i}^{h} = p_{\alpha i} / | \mathbf p_\alpha |$, we arrive at Eq. \eqref {GrindEQ__1_9_}. It should be remembered, however, that in Eq.~\eqref {GrindEQ__1_9_} the values $ X^\mathrm{\omega} \left ( \mathbf x_{\alpha}, t \right) $, $ X_{j}^\mathrm{\omega} \left ( \mathbf x_{\alpha}, t \right) $ do not relate to each other, in general. If necessary, in the three-dimensional case the vector $ X_{j}^\mathrm{\omega} \left ( \mathbf x_{\alpha}, t \right) $ can be considered as two-component in a plane perpendicular to the vector $ e_{\alpha j}^{h} $. The presence of components $ X_{j}^\mathrm{\omega} \left ( \mathbf x_{\alpha}, t \right) $ along $ e_{\alpha j}^{h} $ will not affect the description of processes and phenomena in such systems in any case because of the factor $ \left (\delta_{ij} -e_{\alpha i}^{h} e_{\alpha j}^{h} \right) $ in the right-hand side of Eq.~\eqref {GrindEQ__1_9_} .

It follows from the above that the stochastic effects on the system under consideration in the form of Eq.~\eqref {GrindEQ__1_9_} can be regarded as a generalization of stochastic forces used in the theory of two-dimensional systems of active particles, i.e. ``active fluctuations'' . First, in Ref.\cite{romanczuk2012active} the random forces do not depend on a particle's position in space, whereas Eq. \eqref {GrindEQ__1_9_} allows for the possibility of local influence of stochastic forces on the system. Second, Eq.~\eqref {GrindEQ__1_9_} can be applied both to the two- and three-dimensional systems as well. To see this it is sufficient to consider the expression \eqref {GrindEQ__1_9_} two-dimensional and non-local, put it
\begin{equation} \label{GrindEQ__1_10_}
\begin{gathered}
 \mathbf e_{\alpha }^{h} \equiv  \mathbf e_{h} $, $X^\mathrm{\omega } \left(t\right)\equiv \sqrt{D_{v} } \xi_{v} \left(t\right), \\
 \mathbf X^\mathrm{\omega } \left( \mathbf x_{\alpha } ,t\right)\equiv  \mathbf e_{\phi } \sqrt{D_{\phi } } \xi_{\phi } \left(t\right), \\
 \mathbf e_{h} \mathbf  e_{\phi } =0 , 
\end{gathered}
\end{equation}
where $ e_{h} $ is a unit vector along the direction of motion of a particle, $  \mathbf e_{\phi} $ is a unit vector along the azimuthal angle $ \phi $ and $ D_{\phi} $, $ D_{v} $ are angular and velocity noise intensities, respectively \cite{romanczuk2012active}. Note that in two-dimensional systems, as is known, the  isolated directions can appear in the movement of active particles (so-called head-tail asymmetry). This advantage in the direction of the particles motion is due to the existence of the propulsion mechanism. Thus, due to the head-tail asymmetry in the steady state of the many active particles system it is possible to ``naturally'' fix the reference system by a special choice of the vectors $  \mathbf e_{h} $, $  \mathbf e_{\phi} $. Naturally, the existence of this asymmetry is reflected in the many-particle system characteristics, such as a one-particle distribution function. As it is shown below, the existence of the effects of head-tail asymmetry is also possible in three dimensions, even in the case of a linear friction (see Section 5 of this paper). We emphasize that the source of stochastic effects can be generalized to three dimensions in another, as compared with Eq.~\eqref {GrindEQ__1_9_}, form. Similar to the two-dimensional case, one can use, e.g., the spherical coordinates. However, in this paper it is easier to employ the Cartesian coordinates.

Let us also note the following. The time derivative of the total energy of the system in accordance with the Eqs.~\eqref {GrindEQ__1_1_}, \eqref {GrindEQ__1_6_} is given by
\begin{equation} \label{GrindEQ__1_11_}
\frac{dH}{dt} =-\sum_{1\le \alpha \le N}\frac{ \mathbf p_{\alpha } }{m}  \frac{\partial R}{\partial \mathbf  p_{\alpha } }.
\end{equation}

If we assume that the system has a dissipation due to friction of macroscopic particles, and a regular part of the dissipation function $ R^{r} $ is given by Eq.~\eqref {GrindEQ__1_5_}, then Eq.~\eqref {GrindEQ__1_11_} together with Eq.~\eqref {GrindEQ__1_9_} give
\begin{equation} \label{GrindEQ__1_12_}
\begin{gathered}
\frac{dH}{dt} =-\frac{2}{m} \sum_{1\le \alpha <\beta \le N}R_{\alpha ,\beta }  + \sum_{1\le \alpha \le N}\frac{p_{i\alpha } }{m}  \left[X^\mathrm{\omega } \left( \mathbf x_{\alpha } ,t\right)e_{\alpha i}^{h} \right.\\
\hfill \left. + X_{j}^\mathrm{\omega } \left( \mathbf x_{\alpha } ,t\right)\left(\delta_{ij} -e_{\alpha i}^{h} e_{\alpha j}^{h} \right)\right], 
\end{gathered}
\end{equation}
or
\[\frac{dH}{dt} =-\frac{2}{m} \sum_{1\le \alpha <\beta \le N}R_{\alpha ,\beta }  +\sum_{1\le \alpha \le N}\frac{\left| \mathbf p_{\alpha } \right|}{m} X^\mathrm{\omega } \left( \mathbf x_{\alpha } ,t\right) .\] 

Taking into account that $ \tilde {\gamma} \left ( \mathbf x_{\alpha \beta} \right)> 0 $, see Eq.~\eqref {GrindEQ__1_5_}, in such a system competition between the dissipation due to friction and the pumping of energy from the stochastic field is possible. 

Further task is to obtain the Liouville equation. To this end, for the convenience of further calculations we represent the equations \eqref {GrindEQ__1_6_} in the following form
\begin{equation} \label{GrindEQ__1_13_}
\begin{gathered}
\dot{x}_{\alpha } \left(t\right)=h_{\alpha }^\mathrm{\omega } \left(x_{1} \left(t\right),...,x_{N} \left(t\right)\right), \qquad
1\le \alpha \le N
\end{gathered}
\end{equation}
where we introduce the notation
\begin{equation} \label{GrindEQ__1_14_}
x_{a} \left(t\right)\equiv \left(x_{a} \left(t\right),\;  \mathbf p_{a} \left(t\right)\right).
\end{equation}

Thus, Eq.~\eqref {GrindEQ__1_13_} together with Eqs.~\eqref {GrindEQ__1_14_}, \eqref {GrindEQ__1_5_} reads
\begin{equation} \label{GrindEQ__1_15_}
\begin{gathered}
\dot{ \mathbf x}_{\alpha } \left(t\right)= \mathbf h_{ \mathbf x\alpha }^\mathrm{\omega } \left(x\left(t\right)\right), \\
\dot{ \mathbf p}_{\alpha } \left(t\right)= \mathbf h_{ \mathbf p\alpha }^\mathrm{\omega } \left(x\left(t\right)\right),
\end{gathered}
\end{equation}
where
\begin{equation} \label{GrindEQ__1_16_}
\begin{gathered}
 \mathbf h_{ \mathbf x\alpha }^\mathrm{\omega } \left(x\left(t\right)\right)=\frac{\partial H}{\partial  \mathbf p_{\alpha } }, \\
 \mathbf h_{ \mathbf p\alpha }^\mathrm{\omega } \left(x\left(t\right)\right)=-\frac{\partial H}{\partial  \mathbf x_{\alpha } } -\frac{\partial R}{\partial  \mathbf p_{\alpha } }.
\end{gathered}
\end{equation}

The coordinates and momenta of $ \alpha $-th particle at time $ t $ (see Eq.~\eqref {GrindEQ__1_14_}), are determined by the coordinates and momenta $ x_{0} \equiv \left (x_{1} \left (0 \right), ..., x_{N} \left (0 \right) \right) $ of all the particles at the initial time $ t = 0 $:
\begin{equation} \label{GrindEQ__1_17_}
x_{\alpha }^\mathrm{\omega } \left(t\right)=X_{\alpha }^\mathrm{\omega } \left(t,x_{0} \right)\equiv \left( \mathbf X_{\alpha }^\mathrm{\omega } \left(t,x_{0} \right),\;  \mathbf P_{\alpha }^\mathrm{\omega } \left(t,x_{0} \right)\right),
\end{equation}
where the functions $  \mathbf X_{\alpha}^\mathrm{\omega} \left (t, x_{0} \right) $, $ \mathbf P_{\alpha}^\mathrm{\omega} \left (t, x_{0} \right) $ satisfy the generalized Hamilton equations \eqref {GrindEQ__1_5_} (or equations \eqref {GrindEQ__1_13_}--\eqref {GrindEQ__1_16_}). Introduce
the probability density $ D \left (x_{1} \left (0 \right),\right.$ $\dots, x_{N} \left (0 \right);$ $\left. 0 \right) $ of the initial conditions $ x_{0} \equiv \left (x_{1} \left (0 \right), ..., x_{N} \left (0 \right) \right) $,
\begin{equation} \label{GrindEQ__1_18_}
\int d x_{1} \left(0\right)...dx_{N} \left(0\right)D\left(x_{1} \left(0\right),...,x_{N} \left(0\right);0\right)\equiv \int dx_{0} D\left(x_{0} ;0\right) =1 .
\end{equation}

Then, at time $ t $ the probability density $ D^\mathrm{\omega} \left (x_{1}, ..., x_{N}; t \right) \equiv D^\mathrm{\omega} \left (x; t \right) $, $ x \equiv \left (x_{1}, ..., x_{N} \right) $, ($ N $-particle distribution function) is defined by the expression
\begin{equation} \label{GrindEQ__1_19_}
D^\mathrm{\omega } \left(x_{1} ,...,x_{N} ;t\right)=\int dx_{0} D\left(x_{0} ;0\right) \mathop{\prod }\limits_{1\le \alpha \le N} \delta \left(x_{\alpha } -X_{\alpha }^\mathrm{\omega } \left(t,x_{0} \right)\right).
\end{equation}

In \cite{laskin1998dynamic} a detailed procedure for the obtaining of the Liouville equation for many-particle systems in an external stochastic field neglecting the interaction between the particles is described. In \cite{sliusarenko2015bogolyubov} (see also \cite{laskin1998dynamic}), a similar procedure is used to obtain a generalized Liouville equation for dissipative many-particle systems in the absence of an external stochastic fields.  The $ N $-particle distribution function obeys the continuity equation
\begin{equation} \label{GrindEQ__1_20_}
\frac{\partial D^\mathrm{\omega } }{\partial t} +\sum_{1\le \alpha \le N}\frac{\partial }{\partial x_{\alpha } }  \left(D^\mathrm{\omega } h_{\alpha }^\mathrm{\omega } \right)=0,
\end{equation}
where the function $h_{\alpha }^{\omega } \left(x\left(t\right)\right)$ is given by the expressions \eqref {GrindEQ__1_16_}, \eqref {GrindEQ__1_17_}. This is the Liouville equation generalized to the case of active particles with pair interactions under the influence of external stochastic fields depending on the velocities of the particles. With Eqs. \eqref {GrindEQ__1_13_} -- \eqref {GrindEQ__1_16_} it can be written as:
\begin{equation} \label{GrindEQ__1_21_}
\begin{aligned}
\frac{\partial D^\mathrm{\omega } }{\partial t} +&\sum_{1\le \alpha \le N}\frac{\partial }{\partial  \mathbf x_{\alpha } }  \left(D^\mathrm{\omega } \frac{\partial H}{\partial  \mathbf p_{\alpha } } \right)\\
+&\sum_{1\le \alpha \le N}\frac{\partial }{\partial  \mathbf p_{\alpha } }  \left(D^\mathrm{\omega } \left(-\frac{\partial H}{\partial  \mathbf x_{\alpha } } -\frac{\partial R}{\partial \mathbf  p_{\alpha } } \right)\right)=0.
\end{aligned}
\end{equation}
In what follows we will use the Liouville equation \eqref {GrindEQ__1_21_}, transformed with Eqs.~\eqref {GrindEQ__1_13_} -- \eqref {GrindEQ__1_16_}, \eqref {GrindEQ__1_8_}, and \eqref {GrindEQ__1_9_} to the form
\begin{equation} \label{GrindEQ__1_22_}
\frac{\partial D^\mathrm{\omega } }{\partial t} +\sum_{1\le \alpha \le N}\frac{ \mathbf p_{\alpha } }{m} \frac{\partial D^\mathrm{\omega } }{\partial x_{\alpha } }  +\sum_{1\le \alpha <\beta \le N}\frac{\partial }{\partial  \mathbf p_{\alpha } } D^\mathrm{\omega } \mathbf  F_{\alpha ,\beta }  +\sum_{1\le \alpha \le N}\frac{\partial }{\partial \mathbf  p_{\alpha } } D^\mathrm{\omega } \mathbf  Y_{\alpha }^\mathrm{\omega }  =0,
\end{equation}
where $ \mathbf  F_{\alpha, \beta} $, $ \mathbf  Y_{\alpha}^\mathrm{\omega} $ are determined by Eqs. \eqref {GrindEQ__1_7_} -- \eqref {GrindEQ__1_9_}. Equation ~\eqref {GrindEQ__1_22_} is an example of the evolution equation with multiplicative noise. Now, the goal is to average this equation
over realizations of the external random force $ \mathbf  Y_{\alpha}^\mathrm{\omega} $.

\section{Averaging generalized Liouville equation with Gaussian random force}

We introduce the $ N $ -particle distribution function $ D \left (x_{1}, ..., x_{N}; t \right) $, which is the distribution function $ D ^ {\omega} \left (x_{1}, ..., x_{N}; t \right) $ (see Eq.~\eqref {GrindEQ__1_19_}), averaged over the random external field $ \mathbf Y ^ {\omega} \left ( \mathbf x, t \right) $ with probability density $ W [ \mathbf Y ^ {\omega}] $:

\begin{equation} \label{GrindEQ__2_1_} 
D\left(x_{1} ,...,x_{N} ;t\right)\equiv \left\langle D^\mathrm{\omega } \left(x_{1} ,...,x_{N} ;t\right)\right\rangle_{\omega } ,  \left\langle ...\right\rangle_{\omega } \equiv \int D  \mathbf Y^\mathrm{\omega } \left( \mathbf x,t\right)W[ \mathbf Y^\mathrm{\omega } ]... 
\end{equation} 

Using the averaging operation \eqref {GrindEQ__2_1_} for the equation \eqref {GrindEQ__1_22_}, we obtain:
\begin{equation} \label{GrindEQ__2_2_}
\frac{\partial D}{\partial t} +\sum_{1\le \alpha \le N}\frac{ \mathbf p_{\alpha } }{m} \frac{\partial D}{\partial  \mathbf x_{\alpha } }  +\sum_{1\le \alpha <\beta \le N}\frac{\partial }{\partial  \mathbf p_{\alpha } } D \mathbf F_{\alpha ,\beta }  +\sum_{1\le \alpha \le N}\frac{\partial }{\partial  \mathbf p_{\alpha } } \left\langle D^\mathrm{\omega } \mathbf  Y_{\alpha }^\mathrm{\omega } \right\rangle_{\omega }  =0.
\end{equation}

To have a closed evolution equation for the  distribution function introduced, it is necessary to express the value of $ \left \langle D ^ {\omega}  \mathbf Y_{\alpha} ^ {\omega} \right \rangle_{\omega} $ through $ D \left (x_{1}, ..., x_{N}; t \right) $. We use the so-called Furutsu--Novikov formula \cite{furutsu1963nbs,novikov1964}, which was proved for the case of Gaussian distributions of the external random field. For non-Gaussian 
random fields the Furutsu--Novikov formula is generalized in \cite{peletminskii2003kinetics} (see also \cite{sliusarenko2015bogolyubov}). In this article, we will not recount the latter proof, referring to the works cited above. We use the result of such a proof of \cite{peletminskii2003kinetics} in the case of a Gaussian distribution of multiplicative noise.
Thus, we get 
\begin{equation} \label{GrindEQ__2_3_}
\begin{gathered}
\left\langle Y_{i}^\mathrm{\omega } \left(x_{\alpha } ,t\right)D^\mathrm{\omega } [ \mathbf Y^\mathrm{\omega } ]\right\rangle_{\omega } =Y_{i} \left(x_{\alpha } ,t\right)\left\langle D^\mathrm{\omega } [ \mathbf Y^\mathrm{\omega } ]\right\rangle_{\omega } + \qquad \qquad \qquad \qquad \qquad \\
\hfill \int dx' \int_{-\infty }^{\infty }dt'y_{ij} \left(x_{\alpha } ,x',t-t'\right) \left\langle \frac{\delta D^\mathrm{\omega } [ \mathbf Y^\mathrm{\omega } ]}{\delta Y_{j} \left(x',t'\right)} \right\rangle_{\omega }, 
\end{gathered}
\end{equation}
where $Y_{i} \left(x_{\alpha } ,t\right)\equiv \left\langle Y_{i}^\mathrm{\omega } \left(x_{\alpha } ,t\right)\right\rangle_{\omega } $, $x_{\alpha } \equiv \left\{ \mathbf x_{\alpha } , \mathbf p_{\alpha } \right\}$ and $y_{ij} \left(x_{\alpha } ,x',t-t'\right)$ is a pair correlation function of the external Gaussian noise ($x'\equiv \left\{ \mathbf x', \mathbf p'\right\}$):
\begin{equation} \label{GrindEQ__2_4_}
y_{ij} \left(x_{\alpha } ,x',t-t'\right)=\left\langle Y_{i}^\mathrm{\omega } \left(x_{\alpha } ,t\right)Y_{j}^\mathrm{\omega } \left(x',t'\right)\right\rangle_{\omega } -\left\langle Y_{i}^\mathrm{\omega } \left(x_{\alpha } ,t\right)\right\rangle_{\omega } \left\langle Y_{j}^\mathrm{\omega } \left(x',t'\right)\right\rangle_{\omega } . 
\end{equation}
In what follows we use $ Y_{i} \left (x \right) \equiv 0 $.
Now, let us consider
\begin{equation} \label{GrindEQ__2_5_}
I_{i} \equiv \int d x'\int_{-\infty }^{\infty }dt' y_{ij} \left(x_{\alpha } ,x',t-t'\right)\left\langle \frac{\delta D^\mathrm{\omega } [ \mathbf Y^\mathrm{\omega } ]}{\delta Y_{j}^\mathrm{\omega } \left(x',t'\right)} \right\rangle_{\omega } 
\end{equation}
in more detail. We assume that the pair correlation function $ y_{ij} \left (x_{\alpha}, x ', t-t' \right) $ is different from zero in the interval $ \left | t-t '\right | \le \tau_{0} $. We also assume that when $ t \sim t '$, pair correlation function $ y_{ij} \left (x_{\alpha}, x', t-t '\right) $ has a sharp maximum. Then the functional derivative $ \frac {\delta D ^ {\omega} [ \mathbf Y ^ {\omega}]} {\delta Y_{j} ^ {\omega} \left (x ', t' \right)} $ is to be evaluated only at $ t \approx t '$. Moreover, as shown in \cite{peletminskii2003kinetics,furutsu1963nbs,novikov1964,moiseev1976}, an exact expression for this derivative can be obtained only when $ t \approx t '$.

In fact, the variational derivative $ \frac {\delta D ^ {\omega} [ \mathbf Y ^ {\omega}]} {\delta Y_{j} ^ {\omega} \left (x ', t '\right)} $ at $ t \approx t' $ undergoes a jump:
\begin{equation} \label{GrindEQ__2_6_}
\begin{gathered}
\frac{\delta D^\mathrm{\omega } [ \mathbf Y^\mathrm{\omega } ]}{\delta \delta Y_{j}^\mathrm{\omega } \left(x',t'\right)} \ne 0,\quad t'\le t, \\
\frac{\delta D^\mathrm{\omega } [ \mathbf Y^\mathrm{\omega } ]}{\delta Y_{j}^\mathrm{\omega } \left(x',t'\right)} =0,\quad t'>t.
\end{gathered}
\end{equation}
The latter circumstance is due to the fact that according to the equation \eqref {GrindEQ__1_22_}, the value of $ D ^ {\omega} \left (t \right) $ can not depend on the field $ Y_{j} ^ {\omega} \left ( \mathbf x ', t '\right) $ taken at a later time than $ t $. According to Eq.~\eqref {GrindEQ__2_5_} the integration over $ t '$ in the formula \eqref {GrindEQ__2_5_} is held in the range of $ - \infty $ to $ t $, instead of $ - \infty $ to $ + \infty $.

Differentiating Eq.~\eqref {GrindEQ__1_22_} by $ Y {j} ^ {\omega} \left ( \mathbf x ', t' \right) $ and noting that according to Eq.~\eqref {GrindEQ__2_5_} the derivative $ \frac {\partial} {\partial t} \frac {\delta D ^ {\omega} [ \mathbf Y ^ {\omega}]} {\delta Y_{j} ^ {\omega} \left (x ', t' \right)} $ must have a $ \delta $-like shape in time (while the value $ \frac {\delta D ^ {\omega} [ \mathbf Y ^ {\omega}]} {\delta Y_{j} ^ {\omega} \left (x ', t' \right)} $ does not), the following expression for the functional derivative is obtained (see \cite{peletminskii2003kinetics}):
\begin{equation} \label{GrindEQ__2_7_}
\begin{gathered}
\frac{\delta D^\mathrm{\omega } [ \mathbf Y^\mathrm{\omega } ]}{\delta Y_{j}^\mathrm{\omega } \left(x',t'\right)} \approx -\vartheta \left(t-t'\right)\sum_{1\le \beta \le N}\delta \left(x'-x_{\beta } \right)\frac{\partial D^\mathrm{\omega } [ \mathbf Y^\mathrm{\omega } ]}{\partial p_{\beta j} }, \\
\delta \left(x'-x_{\beta } \right)\equiv \delta \left( \mathbf x'- \mathbf x_{\beta } \right)\delta \left( \mathbf p'- \mathbf p_{\beta } \right)
\end{gathered}
\end{equation}
where $\vartheta \left(t-t'\right)$ is the Heaviside function. This formula allows us to represent $ I_{i} $, Eq.~\eqref {GrindEQ__2_5_}, in the following form (see Eqs.~\eqref {GrindEQ__2_1_}, \eqref {GrindEQ__2_2_}):
\begin{equation} \label{GrindEQ__2_8_}
I_{i} =\int_{-\infty }^{t}dt' \sum_{1\le \beta \le N}y_{ij} \left(x,x_{\beta } ;t-t'\right)\frac{\partial D}{\partial p_{\beta j} }.
\end{equation}
Thus, the averaged Liouville equation, generalized to the case of systems of many particles with active interaction reads
\begin{equation} \label{GrindEQ__2_9_}
\begin{gathered}
\frac{\partial D}{\partial t} +\sum_{1\le \alpha \le N}\frac{ \mathbf p_{\alpha } }{m} \frac{\partial D}{\partial  \mathbf x_{\alpha } }  +\sum_{1\le \alpha <\beta \le N}\frac{\partial D \mathbf F_{\alpha ,\beta } }{\partial  \mathbf p_{\alpha } }  +\sum_{1\le \alpha \le N}\frac{\partial D \mathbf Y\left(x_{\alpha } ,t\right)}{\partial  \mathbf p_{\alpha } } \qquad\\
\hfill - \int_{-\infty }^{t}dt' \sum_{1\le \alpha ,\beta \le N}\frac{\partial }{\partial p_{\alpha i} } y_{ij} \left(x_{\alpha } ,x_{\beta } ;t-t'\right)\frac{\partial D}{\partial p_{\beta j} }  =0.
\end{gathered}
\end{equation}
Taking into account that the pair correlation function $ y_{ij} \left (x_{\alpha}, x_{\beta}; t-t '\right) $ has a sharp maximum at $ t \approx t' $, and also assuming that this function is an even function of difference $t-t'$,
\[
y_{ij} \left (x_{\alpha}, x_{\beta}; t-t '\right) = y_{ij} \left (x_{\alpha}, x_{\beta}; t'-t \right)
\]
then Eq.~\eqref {GrindEQ__2_9_} gets even a simpler form,
\begin{equation} \label{GrindEQ__2_11_}
\begin{gathered}
\frac{\partial D}{\partial t} +\sum_{1\le \alpha \le N}\frac{ \mathbf  \mathbf p_{\alpha } }{m} \frac{\partial D}{\partial  \mathbf x_{\alpha } }  +\sum_{1\le \alpha <\beta \le N}\frac{\partial D \mathbf F_{\alpha ,\beta } }{\partial \mathbf  p_{\alpha } }  \qquad \qquad \qquad \qquad \qquad \\
\hfill - \frac{1}{2} \sum_{1\le \alpha ,\beta \le N}\frac{\partial }{\partial p_{\alpha i} } y_{ij} \left(x_{\alpha } ,x_{\beta } \right)\frac{\partial D}{\partial p_{\beta j} }  =0,
\end{gathered}
\end{equation}
where we introduce the notation:
\begin{equation} \label{GrindEQ__2_12_}
y_{ij} \left(x_{\alpha } ,x_{\beta } \right)\equiv \int_{-\infty }^{\infty }d\tau  y_{ij} \left(x_{\alpha } ,x_{\beta } ;\tau \right).
\end{equation}
Equation \eqref {GrindEQ__2_12_} can be put in another form suitable for further calculations:
\begin{equation} \label{GrindEQ__2_13_}
\begin{gathered}
\frac{\partial D}{\partial t} +\sum_{1\le \alpha \le N}\frac{ \mathbf p_{\alpha } }{m} \frac{\partial D}{\partial \mathbf  x_{\alpha } }  +\sum_{1\le \alpha <\beta \le N}\frac{\partial D \mathbf F_{\alpha ,\beta } }{\partial  \mathbf p_{\alpha } }  \hfill \\
-\frac{1}{2} \sum_{1\le \alpha \le N}\frac{\partial }{\partial p_{\alpha i} } y_{ij} \left(x_{\alpha } ,x_{\alpha } \right)\frac{\partial D}{\partial p_{\alpha j} } - \sum_{1\le \alpha <\beta \le N}\frac{\partial }{\partial p_{\alpha i} } y_{ij} \left(x_{\alpha } ,x_{\beta } \right)\frac{\partial D}{\partial p_{\beta j} }  =0.
\end{gathered}
\end{equation}

Note that, in fact, the developed technique allows to obtain a generalized Liouville equation also
in case of non-Gaussian random field whenever these distributions have moments of any order, see 
\cite{peletminskii2003kinetics}. In the present paper, however, we restrict ourselves to a Gaussian 
external random  field.

For further calculations, we specify the explicit form of the pair correlation function $ y_{ij} \left (x_{\alpha}, x_{\beta} \right) $. Using Eqs. \eqref {GrindEQ__1_9_} and Eq.~\eqref {GrindEQ__2_4_} we arrive at the following expression for $ y_{ij} \left (x_{\alpha}, x_{\beta} \right) $:
\begin{equation} \label{GrindEQ__2_14_}
\begin{gathered}
y_{ij} \left(x_{\alpha } ,x_{\beta } \right)=n_{i\alpha } n_{j\beta } g\left( \mathbf x_{\alpha } , \mathbf x_{\beta } \right)+\left(\delta_{il} -n_{i\alpha } n_{l\alpha } \right)\left(\delta_{jl} -n_{j\beta } n_{l\beta } \right)h\left( \mathbf x_{\alpha } , \mathbf x_{\beta } \right), \\
e_{\alpha i}^{h} =\frac{p_{i\alpha } }{\left| \mathbf p_{\alpha } \right|}, 
\end{gathered}
\end{equation}
where we introduced the notations:
\begin{equation} \label{GrindEQ__2_15_}
\begin{gathered}
g\left( \mathbf x_{\alpha } , \mathbf x_{\beta } \right)\equiv \int_{-\infty }^{\infty }dt \left\langle X^\mathrm{\omega } 
\left( \mathbf x_{\alpha } ,t\right)
X^\mathrm{\omega } \left( \mathbf x_{\beta } ,t'\right)\right\rangle_{\omega },
\\
\delta_{lk} h\left( \mathbf x_{\alpha } , \mathbf x_{\beta } \right)\equiv \int_{-\infty }^{\infty }dt \left\langle X_{l}^\mathrm{\omega } \left( \mathbf x_{\alpha } ,t\right)X_{k}^\mathrm{\omega } \left( \mathbf x_{\beta } ,t'\right)\right\rangle_{\omega }. 
\end{gathered}
\end{equation}
When obtaining expressions \eqref {GrindEQ__2_14_} we assumed that the stochastic force $ Y_{i} ^ {\omega} \left (x, t \right) $ has the following properties:
\begin{equation} \label{GrindEQ__2_16_}
\begin{gathered}
\left\langle X^\mathrm{\omega } \left( \mathbf x_{\alpha } ,t\right)X_{i}^\mathrm{\omega } \left( \mathbf x_{\beta } ,t'\right)\right\rangle =0, \\
\left\langle X^\mathrm{\omega } \left( \mathbf x_{\alpha } ,t\right)\right\rangle =0, \\
\left\langle X_{i}^\mathrm{\omega } \left( \mathbf x_{\beta } ,t'\right)\right\rangle =0 . 
\end{gathered}
\end{equation}
The last two formulas in Eq.~\eqref {GrindEQ__2_16_} are the result of the requirement $ Y_{i} \left (x, t \right) \equiv \left \langle Y_{i} ^ {\omega} \left (x, t \right) \right \rangle_{\omega} = 0 $, see Eq.~\eqref {GrindEQ__1_9_}.

\section{Analogue of the BBGKY chain for systems of identical active particles interacting with external random fields}

Along with the probability density $ D(x_1,...,x_N,t) $ we can introduce the probability of finding one or more particles in the given elements of phase space, regardless of the positions of the remaining 
particles (see also \cite{bogoliubov1959problems,akhiezer1981methods}). These probabilities can be obtained by integrating the function $ D $ over all variables except those that relate to the particles under consideration:
\begin{equation} \label{GrindEQ__3_1_}
\begin{gathered}
f_{S} \left(x_{1} ,...,x_{S} ;t\right)={\cal V}^{S} \int dx_{S+1}  ...\int dx_{N}  D\left(x_{1} ,...,x_{N} ;t\right), \\
x_{\alpha } \equiv \left( \mathbf x_{\alpha } , \mathbf p_{\alpha } \right), 
\end{gathered}
\end{equation}
where $ D \left (x_{1}, ..., x_{N}; t \right) $ satisfies Eq.~\eqref {GrindEQ__2_13_} and $ {\cal V} $ is the system volume. Following the procedure described in \cite{astumian2002brownian,reimann2002brownian}, after some transformations we arrive at the following equation for the $ S $ -particle distribution function $ f_{S} \left (x_{1}, ..., x_{S}; t \right) $:
\begin{equation} \label{GrindEQ__3_2_}
\begin{gathered}
\frac{\partial f_{S} }{\partial t} +\sum_{1\le \alpha \le S}\frac{ \mathbf p_{\alpha } }{m} \frac{\partial f_{S} }{\partial x_{\alpha } }  -\frac{1}{2} \sum_{1\le \alpha \le S}\frac{\partial }{\partial p_{\alpha i} } y_{ij} \left(x_{\alpha } ,x_{\alpha } \right)\frac{\partial f_{S} }{\partial p_{\alpha j} } \qquad \qquad \qquad \\
+\sum_{1\le \alpha <\beta \le S}\frac{\partial f_{S}  \mathbf F_{\alpha ,\beta } }{\partial \mathbf  p_{\alpha } }  -\sum_{1\le \alpha <\beta \le S}\frac{\partial }{\partial p_{\alpha i} } y_{ij} \left(x_{\alpha } ,x_{\beta } \right)\frac{\partial f_{S} }{\partial p_{\beta j} }  =\\
-\frac{1}{\mathfrak{v}} \sum_{1\le \alpha \le S}\frac{\partial }{\partial \mathbf  p_{\alpha } }  \int dx_{S+1}  f_{S+1} \mathbf  F_{\alpha ,S+1} \\
\hfill +\frac{1}{\mathfrak{v}} \sum_{1\le \alpha \le S}\frac{\partial }{\partial p_{\alpha i} } \int dx_{S+1}  y_{ij} \left(x_{\alpha } ,x_{S+1} \right)\frac{\partial f_{S+1} }{\partial p_{S+1j} }, \\
{\mathfrak{v}} \equiv \frac{{\cal V}}{N},
\end{gathered}
\end{equation}
where the quantities $  \mathbf F_{\alpha, \beta} $ and $ y_{ij} \left (x_{\alpha}, x_{\beta} \right) $ are still given by Eqs. \eqref {GrindEQ__1_8_}, \eqref {GrindEQ__2_4_}, \eqref {GrindEQ__2_14_} and \eqref {GrindEQ__2_15_}. As it is easy to see the equation for the $ S $ -particle distribution function includes a $ S + 1 $ -particle distribution function. Thus, in fact, we obtain an infinite chain of kinetic equations \eqref {GrindEQ__3_2_}. These chains are a generalization of the well-known chain of Bogolyubov--Born--Green--Kirkwood--Yvon equations in case of identical active interacting particles under the influence of external stochastic fields. It is necessary to make the following remark. According to the definition \eqref {GrindEQ__3_1_}, the distribution functions of a higher order contain all the information contained in the functions of lower order \cite{bogoliubov1959problems}. This leads to the fact that with the increase in the order $ S $, the distribution functions $ f_{S} \left (x_{1}, ..., x_{S}; t \right) $ are becoming increasingly complex. Since in full description according to Eq.~\eqref {GrindEQ__3_2_} it is necessary to consider the distribution functions up to $ S = N $, we conclude that the resulting chain of equations \eqref {GrindEQ__3_2_} themselves are equivalent to Liouville equation \eqref {GrindEQ__2_13_}. In other words, the most complete description of the studied systems is equally complex both within the framework of the full the distribution function $ D \left (x_{1}, ..., x_{N}; t \right) $, and the one of the many-particle distribution functions $ f_{S} \left (x_{1}, ..., x_{S}; t \right) $.

A significant simplification in description of the state of the system occurs in two cases: when the interaction between the particles is small, or when the number density of particles is small, and the interaction is arbitrary, but is such that does not lead to the formation of bound states \cite{reimann2002brownian}. This simplification in the description is the consequence of the difference in the evolutionary behaviour of a many- and a single-particle distribution functions. In fact, at an early stage of evolution, when the time $ t $ is small compared to the characteristic time of chaotization $ \tau_{0} $, the multi-particle distribution functions $ f_{S} \left (x_{1}, .. ., x_{S}; t \right) $ change rapidly over time, in contrast to the single-particle distribution function $ f_{1} \left (x, t \right) $. Single-particle distribution function experiences significant changes in time at times much longer than the relaxation time of the system $ \tau_{r} $, and $ \tau_{r} \gg \tau_{0} $. Time $ \tau_{0} $, in order of magnitude is determined by the duration of one collision. While time $ \tau_{r} $ in order of magnitude should be the same as the time of the establishment of statistical equilibrium state in the system (for more details see Ref. \cite{akhiezer1981methods}). Such difference in the evolutionary behaviour of the single-particle and many-particle distribution functions formed the basis of the ideas of Bogolyubov about a hierarchy of the system relaxation times \cite{bogoliubov1959problems}. In turn, as mentioned above, based on this idea there have been formulated provisions of the now well-known Bogolyubov--Peletminsky reduced description method for the study of non-equilibrium processes in many-particle systems. The main statements of this method were formulated by N.N.~Bogolyubov for description of the evolution of classical (non-quantum) systems \cite{bogoliubov1959problems}. In case of quantum systems the reduced description method was generalized in works by S.V.~Peletminsky, the most complete quotation of which may be found in \cite{akhiezer1981methods}. We emphasize, however, that in \cite{bogoliubov1959problems,akhiezer1981methods} the systems of many active particles are not considered at all; also they do not deal with the impact of stochastic fields on the many-particle systems.

According to the idea of a hierarchy of relaxation times, the evolution of many-particle system can be divided into several stages. Each subsequent stage of evolution differs from the previous by a simplification in the description of the evolution of system of many particles. 
The simplest scenario for the evolution of systems of many particles is as follows. When $ \tau_{0} \ll t \ll \tau_{r} $ there takes place a kinetic stage of evolution of the system, when the system behaviour can be described by a single-particle distribution function. This description of the system evolution is much easier than that using the multi-particle distribution functions. Further simplification of the description of many-particle systems occurs when $ t \gg \tau_{r} $ (the hydrodynamic stage of evolution of the system), when the behaviour of the system can be described by the hydrodynamic description parameters , for example, the particle number density, the average velocity and the temperature of the medium. Such a gradual simplification of the system description the approaches of the reduced description method are based on \cite{bogoliubov1959problems,akhiezer1981methods}.

In this paper, the method of reduced description of non-equilibrium processes will be used for the derivation of the kinetic equations describing the evolution of systems of interacting active particles in an external random field. The initial equations will be the chain equations \eqref {GrindEQ__3_2_}. 
The mathematical formulation of the idea of a hierarchy of relaxation times of the system is a time-functional dependence of many-particle distribution functions $ f_{S} \left (x_{1}, ..., x_{S}; t \right) $ only through a dependence on time of the parameters of the reduced description at the appropriate stage of evolution. In particular, at the kinetic stage of the evolution the many-particle distribution functions depend on time only through the one-particle distribution function $ f_{1} (x ', t) $:
\begin{equation} \label{GrindEQ__3_3_}
f_{S} \left(x_{1} ,...,x_{S} ;t\right)=f_{S} \left(x_{1} ,...,x_{S} ;f_{1} (x',t)\right)
\end{equation}

In addition to the functional hypothesis \eqref {GrindEQ__3_3_}, the reduced description method is also based on the principle of spatial correlation weakening. In the language of multi-particle distribution functions, this principle can be summarized as follows \cite{akhiezer1981methods}. 
Let $ S $ of the particles can be divided into two sub-groups of particles containing $ S'$ and $ S''$ particles, respectively, $ S = S' + S'' $. If the distance $ R $ between these subgroups of particles increases infinitely, $ R \to \infty $, then due to the weakening of correlations between particles the $ S $ -particle distribution function decomposes into the product of the distribution functions related to the each particles sub-group:
\begin{equation} \label{GrindEQ__3_4_}
f_{S} \left(x_{1} ,...,x_{S} ;t\right)\mathop{\to }\limits_{R\to \infty } f_{S'} \left(x'_{1} ,...,x'_{S} ;t\right)f_{S''} \left(x''_{1} ,...,x''_{S} ;t\right).
\end{equation}

In Eq. \eqref {GrindEQ__3_4_} the sign of ``prime'' is used to indicate the coordinates and momenta of the particles of the subgroup $ S '$, and ``two primes' ' to indicate the coordinates and momenta of the particles of the second subgroup. It should be noted, however, that the principle of spatial correlation weakening Eq.~\eqref {GrindEQ__3_5_} refers to the many-particle distribution functions, for which the thermodynamic limit is made $ N \to \infty $, $ \mathcal{V} \to \infty $, and $ \left (N / \mathcal{V} \right) = const $ \cite{reimann2002brownian}.

According to Eq.~\eqref {GrindEQ__3_3_}, the time derivative of $ \frac {\partial f_{S}} {\partial t} $ in Eq.~\eqref {GrindEQ__3_2_} when $ S \ne 1 $ must be understood as follows:
\begin{equation} \label{GrindEQ__3_5_}
\frac{\partial }{\partial t} f_{S} \left(x_{1} ,...,x_{S} ;f_{1} (x,t)\right)=\int dx'\frac{\delta f_{S} \left(x_{1} ,...,x_{S} ;f_{1} (x,t)\right)}{\delta f_{1} (x',t)}  \frac{\partial f_{1} (x',t)}{\partial t},
\end{equation}
where $ \frac {\delta f_{S} \left (f_{1} (x, t) \right)} {\delta f_{1} (x ', t)} $ is the functional derivative. The single-particle distribution function itself $ f_{1} (x ', t) $ according to Eq.~\eqref {GrindEQ__3_2_} must satisfy the equation: 
\begin{equation} \label{GrindEQ__3_6_}
\frac{\partial f_{1} }{\partial t} +\frac{ \mathbf p_{1} }{m} \frac{\partial f_{1} }{\partial x_{1} } -\frac{1}{2} \frac{\partial }{\partial p_{1i} } y_{ij} \left(x_{1} ,x_{1} \right)\frac{\partial f_{1} }{\partial p_{1j} } =\frac{1}{\mathfrak{v}} L\left(x_{1} ;f_{1} \right),
\end{equation}
where as before $ \mathfrak{v} = {\mathcal{V}} / {N} $ and $ L \left (x_{1}; f_{1} \right) $ is the generalized collision integral defined by the formula
\begin{equation} \label{GrindEQ__3_7_}
\begin{gathered}
L\left(x_{1} ;f_{1} \right)\equiv -\frac{\partial }{\partial  \mathbf p_{1} } \int dx_{2}  f_{2} \left(x_{1} ,x_{2} ;f_{1} \right) \mathbf F_{1,2} \qquad \qquad \qquad \\
\hfill +\frac{\partial }{\partial p_{\alpha i} } \int dx_{2}  y_{ij} \left(x_{1} ,x_{2} \right)\frac{\partial }{\partial p_{2j} } f_{2} \left(x_{1} ,x_{2} ;f_{1} \right).
\end{gathered}
\end{equation}

As is easily seen to close the equation \eqref {GrindEQ__3_6_}, one must obtain the collision integral \eqref {GrindEQ__3_7_} as a functional of particle distribution function for what it is necessary to ``break'' an infinite chain of equations \eqref {GrindEQ__3_2_}. Clearly, this can only be done only in some approximation. In particular, in the system of ``usual'' (non-active) particles such ``break' ' may be implemented in the two cases mentioned above, when the interaction between the particles is small or when the particle density is low, and the interaction is arbitrary but such that does not lead to the formation of bound states \cite{bogoliubov1959problems}. Similar situations can be implemented in the case of a system of identical active particles with interaction, which is discussed in this paper. We will demonstrate this in the case of the weak interaction of all kinds between the active particles and the external noise of low intensity. In other words, we assume the forces $ \mathbf  F_{\alpha, \beta} $ and correlation functions of an external random field are small.

First, however, we make some remarks. Functional relation \eqref {GrindEQ__3_3_} does not necessarily imply an expansion of $ f_{S} \left (x_{1}, ..., x_{S}; f_{1} (x ', t) \right) $ in functional perturbation series by the one-particle distribution function. This expansion must be realized only in one of the above-mentioned cases of chain breaking, namely when the particles density is low. We remind that thus arises the famous question of the possible divergences in higher orders of perturbation theory by a small particle number density and about a renormalization of this theory (see, e.g., \cite{Weinstock1963,kawasaki1965logarithmic,uhlenbeck1963lectures}). In the case of perturbation theory by the weak interaction between the particles, these issues do not appear, as is easily seen from the subsequent calculations (see also \cite{akhiezer1981methods}, \cite{peletminskii2003kinetics}).

\section{Kinetic equations for systems of weakly interacting active particles in external random field of low intensity}

Here we will follow the methodology suggested in \cite{akhiezer1981methods}. Using Eqs.~\eqref {GrindEQ__3_5_}, \eqref {GrindEQ__3_6_}, a chain of equations \eqref {GrindEQ__3_2_} can be written as:
\begin{equation} \label{GrindEQ__4_1_}
-\int dx\frac{\delta f_{S} \left(f_{1} \right)}{\delta f_{1} (x,t)} \frac{ \mathbf p}{m}  \frac{\partial f_{1} (x,t)}{\partial  \mathbf x} +\sum_{1\le \alpha \le S}\frac{ \mathbf p_{\alpha } }{m} \frac{\partial f_{S} \left(f_{1} \right)}{\partial  \mathbf x_{\alpha } } =\frac{1}{\mathfrak{v}} K_{S} \left(f_{1} \right),
\end{equation}
where
\begin{equation} \label{GrindEQ__4_2_}
\begin{gathered}
K_{S} \left(f_{1} \right)\equiv -\mathfrak{v}\sum_{1\le \alpha <\beta \le S}\frac{\partial f_{S}  \mathbf F_{\alpha ,\beta } }{\partial  \mathbf p_{\alpha } }  +\frac{1}{2} \sum_{1\le \alpha \le S}\frac{\partial }{\partial p_{\alpha i} } y_{ij} \left(x_{\alpha } ,x_{\alpha } \right)\frac{\partial f_{S} }{\partial p_{\alpha j} } \qquad \\
+\sum_{1\le \alpha <\beta \le S}\frac{\partial }{\partial p_{\alpha i} } y_{ij} \left(x_{\alpha } ,x_{\beta } \right)\frac{\partial f_{S} }{\partial p_{\beta j} }  - \sum_{1\le \alpha \le S}\frac{\partial }{\partial  \mathbf p_{\alpha } }  \int dx_{S+1}  f_{S+1}  \mathbf  F_{\alpha ,S+1} \\
+\sum_{1\le \alpha \le S}\frac{\partial }{\partial p_{\alpha i} } \int dx_{S+1}  y_{ij} \left(x_{\alpha } ,x_{S+1} \right)\frac{\partial f_{S+1} }{\partial p_{S+1j} }  \\
\hfill -\int dx_{1} \frac{\delta f_{S} \left(f_{1} \right)}{\delta f_{1} (x_{1} ,t)}  \left\{L\left(x_{1} ;f_{1} \right)+\frac{1}{2} \frac{\partial }{\partial p_{1i} } y_{ij} \left(x_{1} ,x_{1} \right)\frac{\partial f_{1} (x_{1} ,t)}{\partial p_{1j} } \right\}.
\end{gathered}
\end{equation}

The chain of equations \eqref {GrindEQ__4_1_}, \eqref {GrindEQ__4_2_} must be supplemented by the ``initial conditions''. To this end, following \cite{bogoliubov1959problems,akhiezer1981methods}, we introduce an auxiliary parameter $ \tau $, having the dimension of time, but does not necessarily representing the physical time. We next consider the many-particle distribution function $ f_ {S} \left ( \mathbf x_ {1} - \frac { \mathbf p_ {1}} {m} \tau,  \mathbf p_ {1}, ..., \mathbf  x_ {S} - \frac { \mathbf p_ {S}} {m} \tau,  \mathbf p_ {S}; f_ {1} \right) $. According to Eq.~\eqref {GrindEQ__3_4_} this function must satisfy the asymptotic relation:
\begin{equation} \label{GrindEQ__4_3_}
f_{S} \left( \mathbf x_{1} -\frac{ \mathbf p_{1} }{m} \tau , \mathbf p_{1} ,..., \mathbf x_{S} -\frac{ \mathbf p_{S} }{m} \tau , \mathbf p_{S} ;f_{1} \right)\mathop{\to }\limits_{\tau \to \infty } \prod_{1\le \alpha \le S}f_{1} \left( \mathbf x_{\alpha } -\frac{ \mathbf p_{\alpha } }{m} \tau , \mathbf p_{\alpha } \right)
\end{equation}

If we define further the shift operator $ \hat {\Lambda} _ {S} ^ {0} $ in the coordinate space with the formula
\begin{equation} \label{GrindEQ__4_4_}
i\hat{\Lambda }_{S}^{0} \equiv \sum_{1\le \alpha \le S}\frac{ \mathbf p_{\alpha } }{m} \frac{\partial }{\partial  \mathbf x_{\alpha } },
\end{equation}
the condition \eqref{GrindEQ__4_3_} may be rewritten as:
\begin{equation} \label{GrindEQ__4_5_}
e^{i\tau \hat{\Lambda }_{S}^{0} } f_{S} \left(\tau \right)\mathop{\to }\limits_{\tau \to \infty } \prod_{1\le \alpha \le S}f_{1} \left(x_{\alpha } \right),
\end{equation}
where $\exp \left(i\tau \hat{\Lambda }_{S}^{0} \right)$ is a so-called ``free evolution operator'' and 
\begin{equation} \label{GrindEQ__4_6_}
\begin{gathered}
f_{S} \left(\tau \right)\equiv f_{S} \left(x_{1} ,...,x_{S} ;e^{-i\tau \hat{\Lambda }_{1}^{0} } f_{1} \left(x'\right)\right) \qquad \quad\\
\hfill =f_{S} \left(x_{1} ,...,x_{S} ;f_{1} \left( \mathbf x'-\frac{ \mathbf p'}{m} \tau , \mathbf p'\right)\right).
\end{gathered}
\end{equation}
Now, Eq.~\eqref{GrindEQ__4_1_} can be written in the following way:
\begin{equation} \label{GrindEQ__4_7_}
\frac{\partial }{\partial \tau } e^{i\tau \hat{\Lambda }_{S}^{0} } f_{S} \left(\tau \right)=\frac{1}{\mathfrak{v}} e^{i\tau \hat{\Lambda }_{S}^{0} } K_{S} \left(\tau \right),
\end{equation}
where
\begin{equation} \label{GrindEQ__4_8_}
\begin{gathered}
K_{S} \left(\tau \right) \equiv K_{S} \left(x_{1} ,...,x_{S} ;e^{-i\tau \hat{\Lambda }_{1}^{0} } f_{1} \left(x'\right)\right) \qquad\\
\hfill =K_{S} \left(x_{1} ,...,x_{S} ;f_{1} \left( \mathbf x'-\frac{ \mathbf p'}{m} \tau , \mathbf p'\right)\right).
\end{gathered}
\end{equation}
Integrating equation \eqref{GrindEQ__4_7_} over $\tau $ within the limits from $-\infty $ to $0$ and using the asymptotic conditions \eqref{GrindEQ__4_5_}, we get
\begin{equation} \label{GrindEQ__4_9_}
f_{S} \left(x_{1} ,...,x_{S} ;f_{1} \left(x'\right)\right)=\prod_{1\le \alpha \le S}f_{1} \left(x_{\alpha } \right) +\frac{1}{\mathfrak{v}} \int_{-\infty }^{0}d\tau  e^{i\tau \hat{\Lambda }_{S}^{0} } K_{S} \left(\tau \right).
\end{equation}
The ratio \eqref {GrindEQ__4_9_} allow to develop a perturbation theory in the weak interaction and the intensity of stochastic effects. Under such assumptions, the value $ K_ {S} \left (\tau \right) $ (see Eq.~\eqref {GrindEQ__4_3_}) can be considered small, and therefore, in the main approximation we have
\[f_{S} \left(x_{1} ,...,x_{S} ;f_{1} \left(x'\right)\right)=\prod_{1\le \alpha \le S}f_{1} \left(x_{\alpha } \right) ,\] 
which implies
\begin{equation} \label{GrindEQ__4_10_}
f_{2} \left(x_{1} ,x_{2} \right)=f_{1} \left(x_{1} \right)f_{1} \left(x_{2} \right).
\end{equation}
Substituting further Eq.~\eqref {GrindEQ__4_10_} into Eq.~\eqref {GrindEQ__3_7_} and using Eqs.~\eqref {GrindEQ__1_7_}, \eqref {GrindEQ__1_8_}, we obtain the following closed kinetic equation:
\begin{equation} \label{4.11a}
\begin{gathered}
\frac{\partial f_{1} \left(x_{1} ,t\right)}{\partial t} +\frac{ \mathbf p_{1} }{m} \frac{\partial f_{1} \left(x_{1} ,t\right)}{\partial  \mathbf  x_{1} } -\frac{1}{2} \frac{\partial }{\partial p_{1i} } y_{ij} \left(x_{1} ,x_{1} \right)\frac{\partial f_{1} \left(x_{1} ,t\right)}{\partial p_{1j} } \qquad\\
= \frac{1}{\mathfrak{v}} \frac{\partial }{\partial  \mathbf p_{1} } f_{1} \left(x_{1} \right)\int dx_{2}  f_{1} \left(x_{2} \right)\left(\frac{\partial V_{1,2} }{\partial  \mathbf x_{1} } +\frac{\partial R_{1,2}}{\partial  \mathbf p_{1} } \right)\\
\hfill+\frac{1}{\mathfrak{v}} \frac{\partial }{\partial p_{1i} } f_{1} \left(x_{1} \right)\int dx_{2}  y_{ij} \left(x_{1} ,x_{2} \right)\frac{\partial f_{1} \left(x_{2} \right)}{\partial p_{2j} },
\end{gathered}
\end{equation}
where the values $ V_ {1,2} $, $ R_ {1,2} ^ {} $ are given by Eqs.~\eqref {GrindEQ__1_2_} -- \eqref {GrindEQ__1_4_} and the correlation function $ y_ {ij} \left (x_ {1 }, x_ {2} \right) $ is still given by Eq.~\eqref {GrindEQ__2_14_}. Equation \eqref{4.11a} and can be rewritten in a slightly different form:
\begin{equation} \label{4.11b}
\begin{gathered}
\frac{\partial f_{1} \left(x_{1} ,t\right)}{\partial t} +\frac{ \mathbf p_{1} }{m} \frac{\partial f_{1} \left(x_{1} ,t\right)}{\partial  \mathbf x_{1} } -\frac{\partial U\left( \mathbf x_{1} ,t\right)}{\partial  \mathbf x_{1} } \frac{\partial f_{1} \left(x_{1} ,t\right)}{\partial  \mathbf p_{1} } \qquad \qquad \qquad\\
= \frac{1}{2} \frac{\partial }{\partial p_{1i} } y_{ij} \left(x_{1} ,x_{1} \right)\frac{\partial f_{1} \left(x_{1} ,t\right)}{\partial p_{1j} } \\
\hfill +\frac{1}{\mathfrak{v}} \frac{\partial }{\partial p_{1i} } f_{1} \left(x_{1} ,t\right)\int dx_{2}  f_{1} \left(x_{2} ,t\right)\left[\frac{\partial R_{1,2} }{\partial p_{1i} } -\frac{\partial y_{ij} \left(x_{1} ,x_{2} \right)}{\partial p_{2j} } \right]
\end{gathered}
\end{equation}
or
\begin{equation} \label{4.11c}
\begin{gathered}
\frac{\partial f_{1} \left(x_{1} ,t\right)}{\partial t} +\frac{ \mathbf p_{1} }{m} \frac{\partial f_{1} \left(x_{1} ,t\right)}{\partial \mathbf  x_{1} } -\frac{\partial U\left( \mathbf x_{1} ,t\right)}{\partial \mathbf  x_{1} } \frac{\partial f_{1} \left(x_{1} ,t\right)}{\partial  \mathbf p_{1} } \qquad \qquad \qquad \\
= \frac{1}{2} \frac{\partial }{\partial p_{1i} } y_{ij} \left(x_{1} ,x_{1} \right)\frac{\partial f_{1} \left(x_{1} ,t\right)}{\partial p_{1j} } \\
\hfill +\frac{1}{\mathfrak{v}} \frac{\partial }{\partial p_{1i} } f_{1} \left(x_{1} ,t\right)\int dx_{2}  \left[R_{1,2} \delta_{i,j} +y_{ij} \left(x_{1} ,x_{2} \right)\right]\frac{\partial f_{1} \left(x_{2} ,t\right)}{\partial p_{2j} } ,
\end{gathered}
\end{equation}
if we consider an average field $U\left( \mathbf x_{1} ,t\right)$, defined by the formula (see Eq.~\eqref {GrindEQ__1_2_}):
\begin{equation} \label{GrindEQ__4_12_}
\begin{gathered}
U\left( \mathbf x_{1} ,t\right)=\frac{1}{\mathfrak{v}} \int d \mathbf x_{2}  V\left( \mathbf x_{1} - \mathbf x_{2} \right)\int d \mathbf p_{2}  f_{1} \left(x_{2} ,t\right),\\
f_{1} \left(x_{2} \right)\equiv f_{1} \left( \mathbf x_{2} , \mathbf p_{2} \right) .
\end{gathered}
\end{equation}

Equations \eqref{4.11a}--\eqref{4.11c} are the kinetic equations for the active particles with pair interactions (potential and ``dissipative'' ones) between the particles  under the influence of active space-dependent fluctuations. We emphasize that all the equations \eqref{4.11a}--\eqref{4.11c} are obtained without using the explicit form of the potential interaction $ V_ {1,2} \equiv V \left ( \mathbf x_ {1} - \mathbf x_ {2} \right) $, dissipation function $ R_ {1 2} $, and the correlation function $ y_ {ij} \left (x_ {1}, x_ {2} \right) $.

Note that the presence of a random force \eqref {GrindEQ__1_9_}, typical for active fluctuations and having a local effect on the particles leads, as it is seen from Eqs.~\eqref {4.11b}, \eqref {4.11c} to an additional interaction between particles, determined by the pair correlation function $ y_ {ij} \left (x_ {1}, x_ {2} \right) $.

\section{Particular cases for spatially homogeneous systems}

Here we demonstrate that the kinetic equations \eqref{4.11a}--\eqref{4.11c} involve known special cases for systems of active particles. To this end, consider a spatially homogeneous state. Then, a single-particle distribution function $ f_ {1} \left ( \mathbf x,  \mathbf p, t \right) $ does not depend on the coordinates,
\begin{equation} \label{GrindEQ__5_1_}
f_{1} \left( \mathbf x, \mathbf p,t\right)\equiv f_{1} \left( \mathbf p,t\right)
\end{equation}
We should specially note that the spatially homogeneous stochastic impact on the system (see Eq.~\eqref{GrindEQ__1_9_}) does not necessarily interdicts with the existence of the states Eq.~\eqref{GrindEQ__5_1_}. The latter are possible in the case of a zero mean of the external random force acting on the system. We remind, that this assumption was made in the present paper beginning from Eq.~\eqref{GrindEQ__2_11_}.
The pair correlation function $ y_ {ij} \left (x_ {1}, x_ {2} \right) $ (see Eqs.~\eqref {GrindEQ__2_14_} -- \eqref {GrindEQ__2_16_}) has the form:
\begin{equation} \label{GrindEQ__5_2_}
y_{ij} \left(x_{1} ,x_{2} \right)=g\left( \mathbf x_{1} - \mathbf x_{2} \right)e_{1i}^{h} e_{2j}^{h} +h\left( \mathbf x_{1} - \mathbf x_{2} \right)\left(\delta_{il} -e_{1i}^{h} e_{1l}^{h} \right)\left(\delta_{jl} -e_{2j}^{h} e_{2l}^{h} \right).
\end{equation}
We recall that according to Eq.~\eqref {GrindEQ__1_4_} all restrictions on the general properties of functions $ R_ {1,2} $ are contained in the expression:
\begin{equation} \label{GrindEQ__5_3_}
R_{1,2} \equiv R\left( \mathbf x_{1} - \mathbf x_{2} , \mathbf p_{1} - \mathbf p_{2} \right),
\end{equation}
which follows from the Galilean invariance of the system in the absence of external influences. Moreover, since the function $ R_ {1,2} $ is a scalar quantity, its dependence on the differences $  \mathbf x_ {1} - \mathbf x_ {2}, \quad  \mathbf p_ {1} - \mathbf p_ {2} $ should be characterized by the expression:
\begin{equation} \label{GrindEQ__5_4_}
R\left( \mathbf x, \mathbf p\right)\equiv R\left( \mathbf x^{2} , \mathbf p^{2} , \mathbf{xp}\right).
\end{equation}

According to Eqs.~\eqref{GrindEQ__5_1_} -- \eqref{GrindEQ__5_4_} equation \eqref{4.11a} transforms into:
\begin{eqnarray}\label{GrindEQ__5_5_}
\nonumber \frac{\partial f_{1} \left( \mathbf p_{1} ,t\right)}{\partial t} &-&\frac{1}{2} \frac{\partial }{\partial p_{1i} } \left[g\left(0\right)e_{1i}^{h} e_{1j}^{h} +h\left(0\right)\left(\delta_{ij} -e_{1i}^{h} e_{1j}^{h} \right)\right]\frac{\partial f_{1} \left( \mathbf p_{1} ,t\right)}{\partial p_{1j} } \\
&=&\frac{\partial }{\partial p_{1i} } f_{1} \left( \mathbf p_{1} ,t\right)\frac{\partial }{\partial p_{1i} } \int d \mathbf p_{2}  f_{1} \left( \mathbf p_{2} ,t\right)\bar{R}\left(\left( \mathbf p_{1} - \mathbf p_{2} \right)^{2} \right)\\
\nonumber &+&\frac{\partial }{\partial p_{1i} } f_{1} \left( \mathbf p_{1} ,t\right)\int d \mathbf p_{2}  \left[\bar{g}e_{1i}^{h} e_{2j}^{h} +\bar{h}\left(\delta_{il} -e_{1i}^{h} e_{1l}^{h} \right)\left(\delta_{jl} -e_{2j}^{h} e_{2l}^{h} \right)\right]\\
\nonumber &\times& \frac{\partial f_{1} \left( \mathbf p_{2} ,t\right)}{\partial p_{2j} } ,
\end{eqnarray}
where we introduce
\begin{eqnarray}\label{GrindEQ__5_6_}
\nonumber \bar{R}\left(\left( \mathbf p_{1} - \mathbf p_{2} \right)^{2} \right) &\equiv& \frac{1}{\mathfrak{v}} \int d \mathbf x R\left( \mathbf x^{2} ,\left( \mathbf p_{1} - \mathbf p_{2} \right)^{2} , \mathbf x\left( \mathbf p_{1} - \mathbf p_{2} \right)\right),\\
\bar{g}&\equiv& \frac{1}{\mathfrak{v}} \int d \mathbf x g\left( \mathbf x\right) ,\\
\nonumber \bar{h}&\equiv& \frac{1}{\mathfrak{v}} \int d \mathbf x h\left( \mathbf x\right) . 
\end{eqnarray}

\paragraph{\textbf{Brownian particles with active fluctuations. Space-independent noise case}}
Here we study quasi-one-dimensional solutions of the kinetic equation \eqref {GrindEQ__5_5_}
in the momentum space,
\begin{equation} \label{GrindEQ__5_7_}
f_{1} \left( \mathbf p,t\right)\equiv f_{1} \left(p,t\right).
\end{equation}
Taking Eqs.~\eqref {GrindEQ__5_6_}, \eqref {GrindEQ__5_7_} it is possible to reduce Eq. \eqref {GrindEQ__5_5_} to the form:
\begin{eqnarray}\label{GrindEQ__5_8_}
\nonumber \frac{\partial f_{1} \left(p_{1} ,t\right)}{\partial t} &=&\frac{\partial }{\partial p_{1i} } e_{1i}^{h} \left\{f_{1} \left(p_{1} ,t\right)\gamma \left(p_{1} ,t\right)p_{1} +\frac{1}{2} g\left(0\right)\frac{\partial f_{1} \left(p_{1} ,t\right)}{\partial p_{1} } \right.\\
&+& \left.\bar{g}f_{1} \left(p_{1} ,t\right)\int d \mathbf p_{2}  \frac{\partial f_{1} \left(p_{2} ,t\right)}{\partial p_{2} } \right\},
\end{eqnarray}
where 
\begin{eqnarray}\label{GrindEQ__5_9_}
\nonumber \gamma \left(p,t\right)\equiv 2\frac{\partial \bar{R}\left(p^{2} ,t\right)}{\partial p^{2} },\\
\bar{R}\left(p^{2} ,t\right)\equiv \int d \mathbf p_{2}  f_{1} \left(p_{2} ,t\right)\bar{R}\left(\left( \mathbf p- \mathbf p_{2} \right)^{2} \right).
\end{eqnarray}
The resulting equation \eqref {GrindEQ__5_8_} is the kinetic equation for active particles with time-dependent non-linear friction (friction factor $ \gamma \left (p, t \right) $). This equation can be regarded as a 
three-dimensional generalization of the kinetic equation for quasi-Brownian particles with active fluctuations, dissipative interaction and space-dependent external stochastic field. This fact may be proven if we make some simplifications of Eq.~\eqref{GrindEQ__5_8_}.

First of all, note, that the term ``quasi-Brownian particles with active fluctuations'' is commonly understood as a system of particles in the presence of friction forces depending on the velocity under the influence of a space-independent
stochastic field given by Eqs.~\eqref {GrindEQ__1_9_} and ~\eqref {GrindEQ__1_10_}, see \cite{romanczuk2012active,lobaskin2013collective}. Consequently, to prove the above assumption, we should pass to the linear friction case in Eq.~\eqref{GrindEQ__5_8_} and refuse the dependence of the external noise on the coordinates. In case of a linear friction the friction coefficient $ \gamma \left (p \right) $ does not depend on the momentum, $ \gamma \left (p \right) \equiv \gamma $, and, according to Eqs.~\eqref {GrindEQ__1_5_}, \eqref {GrindEQ__5_6_}, the value of $ \gamma $ in this case is given by (see \cite{sliusarenko2015bogolyubov}):
\begin{eqnarray}\label{GrindEQ__5_10_}
\nonumber \gamma =\frac{1}{\mathfrak{v}^{2} } \int d \mathbf x\tilde{\gamma }\left( \mathbf x\right) ,\\
\int d \mathbf p f_{1} \left(p,t\right)=\frac{1}{\mathfrak{v}}.
\end{eqnarray}
However, the consequences of the noise space-independence in Eq.~\eqref{GrindEQ__5_8_} are rather hard to see immediately. For this we need to repeat the whole procedure of the kinetic equation derivation until Eqs.~\eqref{4.11a}--\eqref{4.11c}, assuming that the values $X^\omega(\mathbf{x}, t)$, $X^\omega_j(\mathbf{x}, t)$ in Eq.~\eqref {GrindEQ__1_9_} are independent of the coordinates, and the conditions~\eqref{GrindEQ__2_14_}--\eqref{GrindEQ__2_16_} are fulfilled. It turns out, that the result of this procedure is equivalent to equating the value $\bar g$ in Eq.~\eqref{GrindEQ__5_8_} to zero, so that we come to the following equation:
\begin{eqnarray}\label{GrindEQ__5_11_}
\nonumber \frac{\partial f_{1} \left(p_{1} ,t\right)}{\partial t} =\frac{\partial }{\partial p_{1i} } n_{1i} \left\{\gamma p_{1} f_{1} \left(p_{1} ,t\right)+D_{p} \frac{\partial f_{1} \left(p_{1} ,t\right)}{\partial p_{1} } \right\}, \\
g\left(0\right)\equiv 2D_{p}=2m^2D_v. 
\end{eqnarray}
The quantity $g(0) \equiv 2D_p$ is still defined by the relations Eqs.~\eqref{GrindEQ__2_15_} and~\eqref{GrindEQ__5_2_}, keeping in mind the fact that the noise characteristics $X^\omega(\mathbf{x},t)$ does not depend on the coordinate in this case. If in Eq.~\eqref{GrindEQ__5_11_} we pass from the particles' momentum distribution function $f_1(p,t)$ to the distribution function in the velocity, $f_1(v,t)$, $\mathbf{p} = m \mathbf{v}$, then the equation takes the form usual for the case of quasi-Brownian particles with active fluctuations, see, e.g. \cite{romanczuk2012active, grossmann2013self, lobaskin2013collective}. At the same time, the second formula in Eq.~\eqref{GrindEQ__5_11_} connects intensity of the ``momentum'' noise $D_p$ introduced here with intensity of the ``velocity'' noise $D_v$, see Eqs.~\eqref{GrindEQ__1_9_}--\eqref{GrindEQ__1_10_}. This implies that Eq.~\eqref{GrindEQ__5_8_} may be regarded as a kinetic equation for quasi-Brownian particles with active fluctuations, which is generalized for the case of a 3D system with dissipative interaction and a non-local external stochastic field. 

The stationary solution $f_{\infty } \left(p\right)=\mathop{\lim }\limits_{t\to \infty } f_{1} \left(p,t\right)$ of Eq.~\eqref{GrindEQ__5_11_} has a Boltzmann form,
\begin{equation} \label{GrindEQ__5_12_}
f_{\infty } \left(p\right)=A\exp\left(-\frac{\gamma }{2D_{p} } p^{2} \right),
\end{equation}
which is different in a 2D and 3D cases only by the value of the normalizing constant $A$, see Eq.~\eqref{GrindEQ__5_10_}:
\begin{eqnarray}\label{GrindEQ__5_13_}
\nonumber A=\frac{\gamma }{2\pi D_{p} \mathfrak{v}} \quad &\mathrm{for}\quad 2D, \\
A=\frac{1}{\mathfrak{v}} \left(\frac{2\pi D_{p} }{\gamma } \right)^{-3/2} \quad &\mathrm{for} \quad 3D.
\end{eqnarray}
Taking into account normalization \eqref {GrindEQ__5_10_}, \eqref {GrindEQ__5_13_} in the two- dimensional case, the formula \eqref {GrindEQ__5_12_} for stationary distribution function of the active particles coincides with the corresponding expression in Ref. \cite{romanczuk2012active}.

\paragraph{\textbf{Brownian particles with active local fluctuations}}
We now investigate the spatially  homogeneous stationary states of the system under study, in the case of a spatially inhomogeneous external impact. As we already noted, the noise dependence on the coordinates does not exclude the existence of spatially homogeneous states in the system. Let us now consider stationary solution $f_{\infty } \left(p\right)=\mathop{\lim }\limits_{t\to \infty } f_{1} \left(p,t\right)$ of Eq. \eqref{GrindEQ__5_8_}, which is more general than 
Eq. \eqref{GrindEQ__5_11_}. The former in the limit $t\to \infty $ can be written as:
\begin{equation} \label{GrindEQ__5_14_}
f_{\infty } \left(p_{1} \right)\gamma \left(p_{1} \right)p_{1} +\frac{1}{2} g\left(0\right)\frac{\partial f_{\infty } \left(p_{1} \right)}{\partial p_{1} } +\bar{g}f_{\infty } \left(p_{1} \right)\int d \mathbf p_{2}  \frac{\partial f_{\infty } \left(p_{2} \right)}{\partial p_{2} } =0 ,
\end{equation}
where we introduce 
\begin{equation} \label{GrindEQ__5_16_}
\begin{gathered}
g\left(0\right)\equiv 2D_{p} , \\
\tilde{g}\equiv \bar{g}\int d \mathbf p_{2}  \frac{\partial f_{\infty } \left(p_{2} \right)}{\partial p_{2} },
\end{gathered}
\end{equation}
$\bar{g}$ is given by Eq.~\eqref{GrindEQ__5_6_} with Eq.~\eqref{GrindEQ__3_2_}, and
\begin{equation} 
\label{GrindEQ__5_17_} 
\gamma \left(p_{1} \right)=2\mathop{\lim }\limits_{t\to \infty } \frac{\partial \bar{R}\left(p^{2} ,t\right)}{\partial p^{2} } =2\frac{\partial }{\partial p^{2} } \int d \mathbf p_{2}  f_{\infty } \left(p_{2} \right)\bar{R}\left(\left( \mathbf p- \mathbf p_{2} \right)^{2} \right) 
\end{equation} 
defines the non-linear friction forces.
The solution of this equation reads
\begin{equation} \label{GrindEQ__5_15_}
f_{\infty } \left(p\right)\sim \exp \left\{-\frac{1}{D_{p} } \int_{}^{p}dp' \left(\gamma \left(p'\right)p'+\tilde{g}\right)\right\},
\end{equation}
Expressions such as \eqref {GrindEQ__5_15_} are specific to particle systems with non-linear friction under the influence of external active spatially homogeneous fluctuations \cite{romanczuk2011brownian}. It is considered that the non-linear friction is 
responsible for the emergence of head-tail asymmetry \cite{romanczuk2012active,lobaskin2013collective}. 
One should note that the argument of the exponential in Eq.~\eqref{GrindEQ__5_15_} may be positive within one momentum interval, while being negative in another. The argument sign is defined both by the friction (the dependence of quantity $\gamma(p)$ in Eq.~\eqref{GrindEQ__5_15_} on the momentum, see Eq.~\eqref{GrindEQ__5_17_}), and the quantity $\tilde{g}$, which according to Eq.~\eqref{GrindEQ__5_16_} depends on the pair correlation function as is a complex functional of the distribution function itself.
If such intervals of the momentum (or the velocity) are related to a certain (selected) direction, such direction characterizes the head-tail asymmetry. The display of such asymmetry is an emergence of two bell-like peaks stationary distribution functions of the active particles \cite{romanczuk2012active,lobaskin2013collective}. Positions of the maxima of the distribution function (symmetrical with respect to $ p = 0 $) are given by the value of the stationary momentum $ p_ {0} $ of the motion of a ``head'' of the particle. Note that the case $ p_ {0} = 0 $ corresponds to the Boltzmann distribution function, see Eq.~\eqref {GrindEQ__5_12_}.

However, as will be shown below, it follows from the solution \eqref {GrindEQ__5_15_} of equation \eqref {GrindEQ__5_14_} that the stationary distribution function with two maxima (self-propelled particles) can be realized also in the case of a linear friction, namely, when $ \gamma \left (p \right) \equiv \gamma> 0 $, see Eq.~\eqref {GrindEQ__5_10_}. This is due
to the local impact on the system of stochastic forces with active fluctuations. In fact, the general solution in case of linear friction, as is follows from Eq.~\eqref {GrindEQ__5_15_}, is given by:
\begin{equation} \label{GrindEQ__5_18_}
\begin{aligned}
f_{\infty } \left(p\right) \sim {} & \exp \left\{-\frac{\gamma }{2D_{p} } \left(p+\frac{\tilde{g}}{\gamma } \right)^{2} \right\},\\
\tilde{g}\equiv {} & \bar{g}\int d \mathbf p_{2}  \frac{\partial f_{\infty } \left(p_{1} \right)}{\partial p_{2} },\\
\bar{g}\equiv {} & \frac{1}{\mathfrak{v}} \int d \mathbf x g\left( \mathbf x\right).
\end{aligned}
\end{equation}

The display of the head-tail asymmetry is related to the sign of $ \tilde {g} $. Namely, since $ \gamma> 0 $, the positivity of this value, $ \tilde {g}> 0 $, must comply with a purely dissipative case. When $ \tilde {g} <0 $, there are values of momenta, for which the inequality $ \gamma p + \tilde {g} <0 $ is true. For these particles there exists ``propulsion''. In the ``mixed'' case, the single-particle distribution function of active particles has the form 
\cite{romanczuk2012active,lobaskin2013collective}
\begin{equation} \label{GrindEQ__5_19_}
f_{\infty } \left(p\right)=C\left\{\exp \left[-\frac{\gamma }{2D_{p} } \left(p-p_{0} \right)^{2} \right]+\exp \left[-\frac{\gamma }{2D_{p} } \left(p+p_{0} \right)^{2} \right]\right\},
\end{equation}
where $ C $ is the normalization constant. Momentum $ p_ {0} $ in Eq.~\eqref {GrindEQ__5_19_}, characterizing the location of the maxima of the distribution function symmetric with respect to the point $ p = 0 $ is determined by $ \tilde {g} $:
\begin{equation} \label{GrindEQ__5_20_}
p_{0} ={\left|\tilde{g}\right|\mathord{\left/ {\vphantom {\left|\tilde{g}\right| \gamma }} \right. \kern-\nulldelimiterspace} \gamma } .
\end{equation}
The value of $ \tilde {g} $ itself, according to Eqs.~\eqref {GrindEQ__5_16_}, \eqref {GrindEQ__5_18_}, depends on the derivative of the unknown momentum distribution function. Thus, the definition \eqref {GrindEQ__5_18_} with the explicit form of the distribution function \eqref {GrindEQ__5_19_} should be considered as an equation that connects $ \tilde {g} $ to the normalization constant $ C $:
\begin{equation} \label{GrindEQ__5_21_}
\tilde{g}=8\pi \bar{g}C\int_{0}^{\infty }d \mathbf p\frac{\partial }{\partial p}  \left\{\exp \left[-\frac{\gamma }{2D_{p} } \left(p-p_{0} \right)^{2} \right]+\exp \left[-\frac{\gamma }{2D_{p} } \left(p+p_{0} \right)^{2} \right]\right\}
\end{equation}
In turn, the constant $ C $ is determined from the normalization condition (see Eq.~\eqref {GrindEQ__5_10_})
\[\int d \mathbf p f_{\infty } \left(p\right)=\frac{1}{\mathfrak{v}} ,\] 
which can be rewritten after combining with Eq.~\eqref {GrindEQ__5_19_} as
\begin{equation} \label{GrindEQ__5_22_}
\frac{1}{\mathfrak{v}} =C\int d \mathbf p \left\{\exp \left[-\frac{\gamma }{2D_{p} } \left(p-p_{0} \right)^{2} \right]+\exp \left[-\frac{\gamma }{2D_{p} } \left(p+p_{0} \right)^{2} \right]\right\}
\end{equation}
The latter expression is also an equation relating the constant $ C $ and the unknown quantity $ \tilde {g} $. Thus, the equations \eqref {GrindEQ__5_21_} and \eqref {GrindEQ__5_22_} represent a system of two equations to determine two unknown quantities, $ C $ and $ \tilde {g} $, in terms of parameters characterizing the system, namely friction coefficient $ \gamma $ , the number density of particles $ 1 / \mathfrak{v} $ and the parameters of noise with active fluctuations, i.e. the pair correlation function $ \bar {g} $ and $ g \left (0 \right) = 2D_ {p} $, see Eqs.~\eqref {GrindEQ__5_16_}, \eqref {GrindEQ__5_18_}. 
Because of the integration with respect to the total volume in momentum space, equations \eqref {GrindEQ__5_21_}, \eqref {GrindEQ__5_22_} have different forms for two- and three-dimensional cases.

We first consider two-dimensional case. Then Eqs. \eqref {GrindEQ__5_21_} and \eqref {GrindEQ__5_22_} take the form
\begin{equation} \label{GrindEQ__5_23_}
\begin{gathered}
\tilde{g}=-2\pi ^{3/2} \sqrt{\frac{2D_{p} }{\gamma } } \bar{g}C, \\
\frac{1}{\mathfrak{v}} =C\left\{4\pi \frac{D_{p} }{\gamma } +2\pi ^{3/2} p_{0} \sqrt{\frac{2D_{p} }{\gamma } } \mathop{\mathrm{erf}}\left(p_{0} \sqrt{\frac{\gamma }{2D_{p} } } \right)\right\}, p_{0} \equiv \frac{\left|\tilde{g}\right|}{\gamma } ,
\end{gathered}
\end{equation}
where $\mathop{\mathrm{erf}}\left(x\right)$ is the error integral:
\begin{equation} \label{GrindEQ__5_24_}
\mathop{\mathrm{erf}}\left(x\right)\equiv \frac{2}{\sqrt{\pi } } \int_{0}^{x}dy \exp \left(-y^{2} \right).
\end{equation}
In general, Eqs. \eqref {GrindEQ__5_23_} are complex transcendental equations that can be solved numerically. However, in the two extreme cases, namely those of small and large values of the argument $ p_ {0} \sqrt {{\gamma \mathord {\left / {\vphantom {\gamma 2D_ {p}}} \right. \kern- \nulldelimiterspace} 2D_ {p}}} = \sqrt {{\tilde {g} ^ {2} \mathord {\left / {\vphantom {\tilde {g} ^ {2} 2 \gamma D_ { p}}} \right. \kern- \nulldelimiterspace} 2 \gamma D_ {p}}} $ of the error integral \eqref {GrindEQ__5_24_} these equations can be solved analytically. In the case $ p_ {0} \sqrt {{\gamma \mathord {\left / {\vphantom {\gamma 2D_ {p}}} \right. \kern- \nulldelimiterspace} 2D_ {p}}} = \sqrt {{\tilde {g} ^ {2} \mathord {\left / {\vphantom {\tilde {g} ^ {2} 2 \gamma D_ { p}}} \right. \kern- \nulldelimiterspace} 2 \gamma D_ {p}}} \ll 1 $, these solutions are given by
\begin{equation} \label{GrindEQ__5_25_}
\begin{gathered}
C\approx \frac{\gamma }{4\pi D_{p} \mathfrak{v}}, \\
\tilde{g}=-\pi ^{1/2} \frac{\bar{g}}{\mathfrak{v}} \sqrt{\frac{\gamma }{2D_{p} } } , \\
p_{0} \approx \pi ^{1/2} \frac{1}{\mathfrak{v}} \sqrt{\frac{\bar{g}^{2} }{2\gamma D_{p} } } .
\end{gathered}
\end{equation}
With Eqs.\eqref {GrindEQ__5_25_} the inequality above can be written as 
\begin{equation} \label{GrindEQ__5_26_}
\frac{\left|\bar{g}\right|}{D_{p} \mathfrak{v}} \ll 1.
\end{equation}
We have already mentioned that the presence of ``head-tail'' asymmetry 
depends on the sign of $ \tilde {g} $. According to the analysis above it can be concluded from Eq.~\eqref {GrindEQ__5_25_}, when $ \bar {g} <0 $, $ \tilde {g}> 0 $ , the considered system of active particles does not have self-propelled property. Then, according to Eq.~\eqref {GrindEQ__5_18_} only the shift of the maximum of the distribution function defined by the formula \eqref {GrindEQ__5_25_} is observed. If $ \bar {g}> 0 $, then $ \tilde {g} <0 $ and the case of ``head-tail'' asymmetry with a two bell-like peaks distribution function with parameters defined by Eq. \eqref {GrindEQ__5_25_} is realized.

Now consider the opposite case $p_{0} \sqrt{\gamma / 2D_{p} } =\sqrt{\tilde{g}^{2} / 2\gamma D_{p} }  \gg 1$. Then, in the main order we get from Eq. \eqref {GrindEQ__5_23_}:
\begin{equation} \label{GrindEQ__5_27_}
\begin{gathered}
C\approx \frac{1}{4\pi ^{3/2} \mathfrak{v}} \frac{\gamma }{D_{p} } \sqrt{\frac{2\gamma D_{p} }{\left|\tilde{g}\right|^{2} } } , \\
\tilde{g}=-2\pi ^{3/2} \sqrt{\frac{2D_{p} }{\gamma } } \bar{g}C, \\
p_{0} =\sqrt{\frac{\left|\bar{g}\right|}{\gamma \mathfrak{v}} } 
\end{gathered}
\end{equation}
From here one can see that $ \tilde {g}> 0 $ if $ \bar {g} <0 $. In this case the maximum of the distribution function is shifted according
to Eq. \eqref {GrindEQ__5_18_}. In the opposite case, $ \bar {g}> 0 $ the value $ \tilde {g} $ is negative, $ \tilde {g} <0 $, and the particle distribution function is defined by Eqs. \eqref {GrindEQ__5_19_}, \eqref {GrindEQ__5_27_}. We also add that the inequality $ p_ {0} \sqrt {{\gamma \mathord {\left / {\vphantom {\gamma 2D_ {p}}} \right. \kern- \nulldelimiterspace} 2D_ {p}}} = \sqrt {{\tilde {g} ^ {2} \mathord {\left / {\vphantom {\tilde {g} ^ {2} 2 \gamma D_ { p}}} \right. \kern- \nulldelimiterspace} 2 \gamma D_ {p}}} \gg 1 $, with the use of Eq. \eqref {GrindEQ__5_27_}
can be transformed into the relation
\begin{equation} \label{GrindEQ__5_28_}
\frac{\left|\bar{g}\right|}{D_{p} \mathfrak{v}} \gg 1,
\end{equation}
which is the opposite to Eq. \eqref{GrindEQ__5_26_}. 

Now let us return to Eqs. \eqref {GrindEQ__5_21_} and \eqref {GrindEQ__5_22_}, and study their solution in case of three-dimensional system of active particles with linear friction and active space-dependent (i.e., local in space) fluctuations. In this case, Eqs. \eqref {GrindEQ__5_21_}, \eqref {GrindEQ__5_22_} are tranformed to a form substantially different from Eq.~\eqref {GrindEQ__5_23_}:
\begin{equation} \label{GrindEQ__5_29_}
\begin{gathered}
C=\frac{1}{2v} \left(\frac{\pi \gamma }{2D_{p} } \right)^{3/2} , \\
\tilde{g}+2\frac{\left|\tilde{g}\right|}{\gamma } \frac{\pi ^{3} \bar{g}}{\mathfrak{v}} \frac{\gamma }{D_{p} } \mathop{\mathrm{erf}}\left(\sqrt{\frac{\tilde{g}^{2} }{2\gamma D_{p} } } \right)=-8\frac{\pi \bar{g}D_{p} }{\gamma \mathfrak{v}} \left(\frac{\pi \gamma }{2D_{p} } \right)^{3/2} , \\
p_{0} \equiv \frac{\left|\tilde{g}\right|}{\gamma } ,
\end{gathered}
\end{equation}
where $\mathop{\mathrm{erf}}\left(x\right)$ is still given by Eq.~\eqref {GrindEQ__5_24_}. As in the previous case of the two-dimensional system of active particles, the second equation in Eq.~\eqref {GrindEQ__5_29_} in general form can be solved only numerically. However, in the two extreme cases examined above, this equation can be solved analytically. Namely, in the case of small values of the parameter $p_{0} \sqrt{{\gamma \mathord{\left/ {\vphantom {\gamma  2D_{p} }} \right. \kern-\nulldelimiterspace} 2D_{p} } } =\sqrt{{\tilde{g}^{2} \mathord{\left/ {\vphantom {\tilde{g}^{2}  2\gamma D_{p} }} \right. \kern-\nulldelimiterspace} 2\gamma D_{p} } } \ll 1$ the solution of
Eq. \eqref {GrindEQ__5_29_} is determined by
\begin{equation} \label{GrindEQ__5_30_}
\begin{gathered}
\tilde{g}\approx -4\pi ^{2} \frac{\bar{g}}{\mathfrak{v}} \left(\frac{\pi \gamma }{2D_{p} } \right)^{1/2}, \\
p_{0} \approx 4\pi ^{2} \frac{\left|\bar{g}\right|}{vD_{p} } \left(\frac{\pi D_{p} }{2\gamma } \right)^{1/2},
\end{gathered}
\end{equation}
and, as is easily seen directly, the ratio $p_{0} \sqrt{{\gamma \mathord{\left/ {\vphantom {\gamma  2D_{p} }} \right. \kern-\nulldelimiterspace} 2D_{p} } } =\sqrt{{\tilde{g}^{2} \mathord{\left/ {\vphantom {\tilde{g}^{2}  2\gamma D_{p} }} \right. \kern-\nulldelimiterspace} 2\gamma D_{p} } } \ll 1$ can be reduced to the form Eq.~\eqref {GrindEQ__5_26_}, valid for the two-dimensional case. We also note that similar to the two-dimensional case, in a three dimensional system of many active particles at $ \bar {g}> 0 $ we have $ \tilde {g} <0 $, which indicates the possibility of a ``head-tail'' asymmetry. In turn, when $ \bar {g} <0 $ 
we have $ \tilde {g}> 0 $, and the stationary state is characterized by the distribution function with a single maximum shifted to the right by the value $ p_ {0} $, see Eqs.~\eqref {GrindEQ__5_18_}, \eqref {GrindEQ__5_20_}.

At large values of the parameter $p_{0} \sqrt{{\gamma \mathord{\left/ {\vphantom {\gamma  2D_{p} }} \right. \kern-\nulldelimiterspace} 2D_{p} } } =\sqrt{{\tilde{g}^{2} \mathord{\left/ {\vphantom {\tilde{g}^{2}  2\gamma D_{p} }} \right. \kern-\nulldelimiterspace} 2\gamma D_{p} } } $, $p_{0} \sqrt{{\gamma \mathord{\left/ {\vphantom {\gamma  2D_{p} }} \right. \kern-\nulldelimiterspace} 2D_{p} } } =\sqrt{{\tilde{g}^{2} \mathord{\left/ {\vphantom {\tilde{g}^{2}  2\gamma D_{p} }} \right. \kern-\nulldelimiterspace} 2\gamma D_{p} } } \gg 1$ the solution of equation \eqref{GrindEQ__5_29_} is given by:
\begin{equation} \label{GrindEQ__5_31_}
\begin{gathered}
\tilde{g}\approx -8\frac{\pi \bar{g}}{\mathfrak{v}} \left(\frac{\pi \gamma }{2D_{p} } \right)^{3/2} \frac{D_{p} }{\gamma } \left(1-2\pi ^{3} \frac{\bar{g}}{vD_{p} } \right)^{-1} , \\
p_{0} \approx 4\pi ^{2} \frac{\bar{g}}{vD_{p} } \sqrt{\frac{\pi D_{p} }{2\gamma } } \left|1-2\pi ^{3} \frac{\bar{g}}{vD_{p} } \right|^{-1}.
\end{gathered}
\end{equation}
Analysing Eq.~\eqref{GrindEQ__5_31_} we find that at $\bar{g}>0$ in the domain of large values of the parameter the negative $\tilde{g}$ is only possible if $2\pi ^{3} \frac{\bar{g}}{vD_{p} } \mathop{<}\limits_{\sim } 1$. In this case the expression for $p_{0} $ can be simplified,
\begin{equation} \label{GrindEQ__5_32_}
p_{0} \approx \sqrt{\frac{2D_{p} }{\pi \gamma } } \left|1-2\pi ^{3} \frac{\bar{g}}{vD_{p} } \right|^{-1} ,
\end{equation}
and the condition $\sqrt{{\tilde{g}^{2} \mathord{\left/ {\vphantom {\tilde{g}^{2}  2\gamma D_{p} }} \right. \kern-\nulldelimiterspace} 2\gamma D_{p} } } \gg 1$ can be replaced with an equivalent one,
\begin{equation} \label{GrindEQ__5_33_}
0<\frac{\bar{g}}{vD_{p} } \mathop{<}\limits_{\sim } \frac{1}{2\pi ^{3} }. 
\end{equation}
Thus, we conclude that self-propelled particles can be realized in the case of large values of the parameter $\sqrt{{\tilde{g}^{2} \mathord{\left/ {\vphantom {\tilde{g}^{2}  2\gamma D_{p} }} \right. \kern-\nulldelimiterspace} 2\gamma D_{p} } } $ and in three-dimensional system, but the criteria \eqref {GrindEQ__5_33_} in this motion differ significantly from that in the two-dimensional case, see Eq.~\eqref {GrindEQ__5_28_}.

It should be noted that if $ \bar {g} = 0 $ the characteristic momentum $ p_ {0} $ is always zero, both in two- and three-dimensional cases, see Eq.~\eqref {GrindEQ__5_20_} and the original equation \eqref {GrindEQ__5_14_}. This should have been expected, since this case corresponds to the degeneration of the two bell-like peaks single-particle distribution function into a symmetrical one (with respect to $ p = 0 $) of the Gaussian type, with parameters that coincide with those of Eqs.~\eqref {GrindEQ__5_12_}, \eqref {GrindEQ__5_13_}.

\section{Conclusion}

In this paper we propose a microscopic approach to the construction of the kinetic theory of many-particle systems with dissipative and potential interactions in the presence of active fluctuations. The approach is based on a generalization of Bogolyubov--Peletminsky reduced description method applied to the systems of many active particles. It is shown that in the framework of the microscopic approach is possible to construct the kinetic theory of active particles both in the case of two-dimensional, and three-dimensional systems, the availability of non-linear friction (dissipative interaction), as well as local nature of an external random field interaction with active fluctuations.We obtained general kinetic equations for these systems in the case of a weak interaction between the particles (both potential and dissipative) and low-intensity active fluctuations. We define some particular cases in which the derived kinetic equations have solutions that match with the results for the systems of active particles known from earlier works by other authors. It is also shown that one of the consequences of the local nature of the active fluctuations is a manifestation of a head-tail asymmetry and a self-propelling, typical for systems of active particles, even in the case of a linear friction, see (\ref{GrindEQ__5_23_}--\ref{GrindEQ__5_33_}).

We remind in this context that formulas \eqref {GrindEQ__5_23_} -- \eqref {GrindEQ__5_33_} describe only two special limiting cases of the existence of two-dimensional and three-dimensional systems with the ``self-propelled'' particles. The appearance of the obtained expressions coincide with the one of the analogous expressions, see, for example, \cite{romanczuk2012active,lobaskin2013collective}. 
However, we need to note that the mentioned papers do not deal with three-dimensional cases. 
 However, in this article the nature of the phenomenon of ``self-propelling'' is associated with a local (individual) exposure to the particles of the external stochastic field with active fluctuations, see Eq.~\eqref {GrindEQ__1_9_}. Besides, the parameters of the self-propelled motion are self-consistently expressed by the internal characteristics of many-body system - the number density of particles in the system, the parameters of the dissipative function, and characteristics of the external influence - the pair correlation function of the active random field. Note that the stationary direction of ``head-tail'' asymmetry within the spatially homogeneous model (see Eqs.~\eqref {GrindEQ__5_7_}, \eqref {GrindEQ__5_8_}) can not be determined. To define it, we should introduce an interaction, even an arbitrarily small, but violating the spatial uniformity of the problem. In this sense, the situation expressed by Eqs. \eqref {GrindEQ__5_19_} -- \eqref {GrindEQ__5_28_}, is similar to the situation with a phase transition to the magnetic ordering in ferromagnets, see, e.g., \cite{akhiezer1968spin}. As it is known, the value of the total magnetic momentum in a ferromagnet in the main approximation is determined by the isotropic exchange interaction. The direction of the magnetization is at the same time given by non-isotropic weak relativistic interactions.

In this regard, we note that the kinetic equations \eqref {4.11a} -- \eqref {4.11c} are general in the sense that they describe quite a number of different may-particle systems, both two-dimensional and three-dimensional ones, with active local fluctuations, space homogeneous and inhomogeneous as well, including different variations of non-linear friction. However, the study of various particular cases of solutions of kinetic equations \eqref {4.11a} -- \eqref {4.11c} is beyond the scope material of this paper. As outlined above, the main objective of this work is the development of microscopic approach to the derivation of the general kinetic equations for active particles with nonlinear friction under the influence of active fluctuations, including a generalization to the case of the three-dimensional systems.

We also note that the suggested microscopic approach to the construction of the kinetic theory of many-particle systems with dissipative interaction and active fluctuations allow further generalization. It can be generalized, in particular, to the case of simultaneous presence of both active and passive fluctuations. Furthermore, a non-Gaussian stochastic nature of external effects can be taken into account.

\section*{References}

\bibliography{RDMbibfile}

\end{document}